\documentclass[journal,10pt]{IEEEtran}
\usepackage[english]{babel}
\usepackage[utf8]{inputenc}
\usepackage{graphicx}      
\usepackage{amsmath,color}
\usepackage{amsfonts}
\usepackage{amssymb}
\usepackage{fancyhdr}
\usepackage{float} 
\usepackage{multirow}

\usepackage{stfloats}
\usepackage{cite}

\usepackage[dvipsnames]{xcolor}

\begin{document}

\begin{titlepage}
{\bfseries\LARGE IEEE Copyright Notice\par}
\vspace{1cm}
{\scshape\Large © 2020 IEEE.  Personal use of this material is permitted.  Permission from IEEE must be obtained for all other uses, in any current or future media, including reprinting/republishing this material for advertising or promotional purposes, creating new collective works, for resale or redistribution to servers or lists, or reuse of any copyrighted component of this work in other works. \par}
\vspace{1cm}
{\Large This is the author's version of an article that has been published in this journal. Changes were made to this version by the publisher prior to publication. \par}
\vspace{1cm}
{\Large The final version of record is available at http://dx.doi.org/10.1109/TVT.2020.3008336 \par}
\end{titlepage}

\title{Distributed Utility Optimization in Vehicular Communication Systems}

\author{Miguel A. Diaz-Ibarra, Daniel U. Campos-Delgado,~\IEEEmembership{Senior Member,~IEEE}, Carlos A. Gutierrez,~\IEEEmembership{Senior Member,~IEEE}, Jose M. Luna-Rivera,~\IEEEmembership{Member,~IEEE}, Francisco J. Cabrera Almeida~\IEEEmembership{Member,~IEEE}

\thanks{M.A. Diaz-Ibarra, D.U. Campos-Delgado, C.A. Gutierrez and J.M. Luna-Rivera are with Faculty of Sciences, Universidad Autonoma de San Luis Potosi, Av. Chapultepec No. 1570, San Luis Potosi, MX, email: diazibarramiguelangel@gmail.com, ducd@fciencias.uaslp.mx, mlr@fciencias.uaslp.mx, cagutierrez@ieee.org.
F.J. Cabrera Almeida is with Universidad de Las Palmas en Gran Canaria, Instituto Universitario para el Desarrollo Tecnológico y la Innovación en Comunicaciones, Spain, email: francisco.cabrera@ulpgc.es.} 
\thanks{Manuscript received XXX, XX, 20XX; revised XXX, XX, 20XX.}}

\markboth{IEEE Transactions on Vehicular Technology,~Vol.~XX, No.~XX, XXX~20XX}
{}

\maketitle

\begin{abstract}
In this paper, we study the problem of utility maximization in the uplink of vehicle-to-infrastructure communication systems. The studied  scenarios consider four practical aspects of mobile radio communication links:  i) Interference between adjacent channels, ii) interference between roadside units along the way, iii) fast and slow channel fadings, and iv) Doppler shift effects. We present first the system model  {for the IEEE 802.11p standard}, which considers a communication network between vehicles and roadside infrastructure. Next, we formulate the problem of utility maximization in the network, and propose a  {distributed} optimization scheme. This  {distributed} scheme is based on a two-loop feedback configuration, where an outer-loop establishes the optimal signal to interference-noise ratio (SINR) that maximizes the utility function per vehicle and defines a quality-of-service objective. Meanwhile, inner-control loops adjust the transmission power to achieve this optimal SINR reference in each vehicle node regardless of interference, time-varying channel profiles and network latency.  {The computation complexity of the distributed utility maximization scheme is analyzed for each feedback loop.} Simulation results indicate that the proposed scheme reaches the objective SINRs that maximize utility and improve energy efficiency in the network  {with a low time cost}. The results also show that the maximum utility is consistently achieved for different propagation scenarios inside the vehicular communication network.
\end{abstract}


\IEEEpeerreviewmaketitle

\section{Introduction}

\subsection{ {Motivation}}

Vehicular communication networks (VCN) are an emerging technology aimed at improving road safety and traffic management \cite{Vehicular_Networking, DSRC,Vehicular_Communications}. These networks will enable vehicles to share their driving information with other vehicles, and also with the infrastructure installed along the road in real-time. Two types of messages will be employed to transmit (broadcast) such an information: periodic messages (also called beacons), and event-driven messages \cite{Distributed_Fair_Transmit}. The periodic messages are aimed to preserve the vehicles' safety through the dissemination of non-critical information, e.g., updates about road conditions and other vehicles' status (e.g., current position, speed and direction of motion). In turn, event-driven messages are intended for the notification of potentially dangerous events arising randomly, such as an emergency vehicle approaching at high speed, or the sudden braking of a vehicle ahead on the road. Thereby, drivers will be warned in advance to make decisions that avoid accidents, or that contribute to keep a proper traffic flow. The regulation of the VCN is currently driven by two main standards: The IEEE 802.11 Standard for dedicated short-range communications (DSRC) in vehicular environments, and the Long-Term Evolution (LTE) Standard for fourth-generation (4G) mobile cellular communications. The vehicular component of this latter standard is commonly referred to as the LTE-V, or as the LTE-V2X \cite{molina2018ieee}. 

\subsection{State of the Art}

Regardless of the underlying standard, a main concern in VCN is to meet and maintain the particular requirements of quality of service (QoS) of  each vehicle. Power control techniques provide effective solutions to the problem of achieving QoS in wireless communication networks whose performance is affected by inter-user interference, and by the effects of multipath mobile radio channels \cite{Chiang}. Different approaches have been considered in the literature for power control in VCN. In \cite {PowerControlForBroadcastV2V}, two power control algorithms were proposed to mitigate adjacent channel interference (ACI) in vehicle-to-vehicle (V2V) communications. A distributed transmit power control method is suggested in \cite{Vehicle-to-Vehicle} to regulate the load of periodic messages, and to guarantee high priority of event-driven  messages. In \cite{Power-Aware}, a technique called power-aware link quality estimation (PoLiQ) was introduced to estimate the quality of the links in VCN, which is based on the periodic reception of packages from neighboring nodes, and on the transmission parameters of the vehicles. In \cite {Power_Control_in_D2D-based_Vehicular_2015}, the authors study how to efficiently apply the concept of device-to-device (D2D) communications on a cellular network to support V2V connections (called D2D-V).  Strategies to reuse channel selection and optimal power control are considered in that paper for maximizing sum rate and the minimum achievable data transmission rate. Meanwhile, spectrum sharing and power allocation for D2D-enabled VCN is investigated in \cite{Liang}, where channel uncertainty, maximal sum capacity, and reliability of the V2V links were considered. Alternated and distributed optimization approaches for rate and power control are studied in \cite {Distributed_rate_and_power_control} for utility maximization in DSRC systems.

 {The concept of \textit{utility} is originally related to economics and game theory \cite{Luong}. Utility establishes a measure of value that an individual receives from some service. In wireless communications, the utility could be related to different resources:  flow rate, energy efficiency, secrecy capacities, download delay, etc. \cite{Luong,Kons2017, Chen2018, Zheng2014, Zhang2019, Gao2018}. In fact, for some utility functions as flow rate and energy efficiency, there is a direct  relation to the transmission power, which is the optimization variable in this work.} So far, the solutions to the resource optimization problem for VCN have focused mostly on the LTE-V standard. 

\subsection{ {Contributions}}

To the best of the authors' knowledge, there are no studies dealing with the problem of power control for utility maximization in VCN based on the IEEE 802.11p standard. In this paper, we aim to close the gap through the following original contributions: 
\begin{itemize}
	\item We formulate the problem of utility maximization in vehicle-to-infrastructure (V2I) communication systems based on the IEEE 802.11 Standard. Our formulation considers four practical aspects in the V2I communication links: i) Interference between adjacent channels in the IEEE 802.11p standard, ii) interference among roadside units (RSUs) along the road, iii) fast and slow channel fadings, and iv) Doppler shift effect.
	\item To solve the aforementioned problem, we derive a practical  {distributed} optimization scheme by considering a two-loop feedback structure at different time-scales.  {A network central unit implements an outer-loop that updates regularly the objective SINRs to the RSUs to maximize the network utility}. Meanwhile, inner-control loops adjust the transmission power of each on-board unit (OBU) to achieve the desired SINRs despite channel variations and network latency. As a result, each OBU has a time-varying SINR reference to maximize overall network utility.
	\item  {We analyze the computation complexity of the outer and inner-loops in the utility maximization methodology. One key advantage of our proposal is that the complexity is distributed among the network central unit, RSUs and OBUs to facilitate a real-time implementation.}
	\item We present a comparative analysis of the network utility considering three propagation scenarios and different mobility profiles of the vehicles: a) some OBUs approach and others drive away from the RSU, b) all the OBUs approach to the RSU, and c) all the OBUs move away from it. We compare the resulting network utility to fixed target SINR values in the inner-control loops.
\end{itemize}

\begin{figure*}[t]
	\centering
	\includegraphics[width=0.7\linewidth]{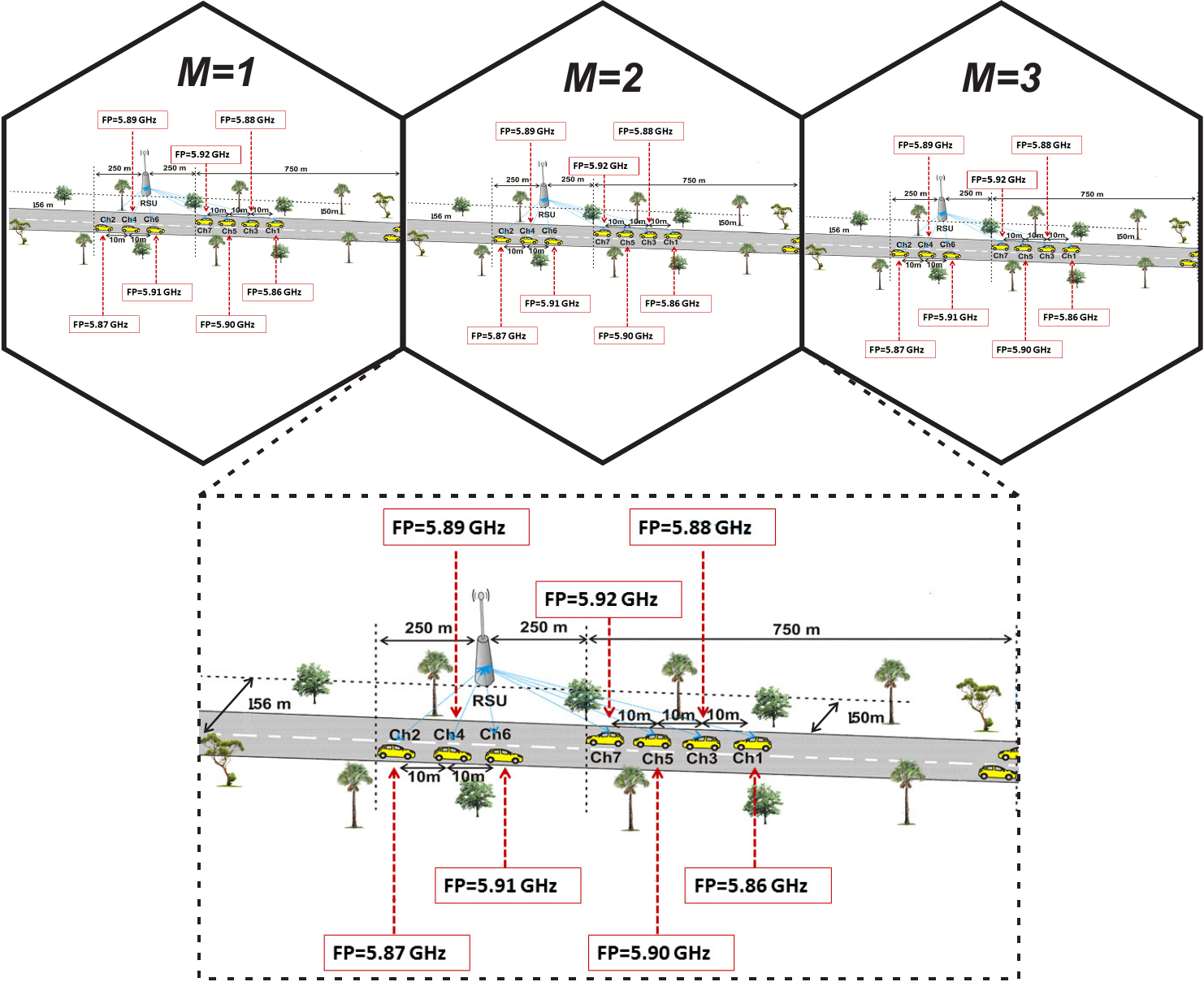}
	\caption{Structure of the  studied vehicular communication network with RSUs and OBUs.}
	\label{Modelo}
\end{figure*}

The rest of the paper is organized as follows. The  {network} system model of a V2I communication system based on the IEEE 802.11p standard with $M$ RSUs and $M \cdot U$ OBUs is described in Section II. The problem of utility maximization is  addressed in Section III. First, the network utility concept is described, and then  {a distributed} strategy for the network utility optimization is presented.  {In addition, Section III introduces the power allocation scheme implemented through inner-control loops to achieve the desired QoS, and also the computation complexity of the proposed scheme is analyzed}. A numerical performance evaluation of the proposed utility maximization strategy is described in Section IV for different propagation patterns of the OBUs in the network. Finally, Section V presents the conclusions.

\section{System Model}

In this paper, we consider a V2I communications network based on the IEEE 802.11p standard, as the one shown in Fig.~\ref{Modelo}. The network comprises two types of nodes: OBUs, which are vehicles with DSRC capabilities; and RSUs, which are static DSRC devices installed on the road, e.g., a base station (BS). We address the problem of resource optimization for the uplink of this V2I communication network (i.e. OBUs to RSU), considering a two-lane highway scenario with $M$ RSUs and $M \cdot U$ OBUs (see Fig. \ref{Modelo}). We assume that the RSUs have the same configuration, and each OBU is assigned to a frequency channel of 10 MHz in the 5.9 GHz band.  {We focus on the uplink of a V2I communication network, since as will be discussed next, efficient resources allocation is crucial to reduce the interference between adjacent channels, and the interference from RSU to RSU. Nonetheless, the  proposed resources optimization methodology could be also adapted to downlink communications.}

\subsection{ {Interference Sources in IEEE 802.11p}}

We assume that the main factor limiting the performance of the V2I communication network is interference, which can be quantified by the SINR \cite{PowerControlForBroadcastV2V}. Two major types of interference can be identified for VCN  {in the IEEE 802.11p standard}:
\begin{itemize}
	\item  {\textit{Adjacent channel interference (ACI)} is the result of having multiple adjacent frequency channels, which are active simultaneously, as shown in Fig.  \ref{Masks}. This type of interference is caused by nonlinear effects of the transmitter's filters \cite{PowerControlForBroadcastV2V}.}	
	\item  {\textit{Interference between RSUs} is caused by using the same frequency channels for different nearby RSUs}.
\end{itemize}
To analyze the effects that such forms of interference have on the network's resources, we will consider the specifications issued by the United States Federal Communications Commission (FCC) concerning the electromagnetic spectrum for DSRC applications. In the 1990s, the FCC allocated a bandwidth of 75 MHz in the 5.9 GHz band for DSRC systems. This bandwidth is divided into seven 10 MHz channels, one reserved as a control channel and six considered for communication services. The service channels are used for infotainment (information and entertainment) applications and traffic management, while the control channel is reserved for road safety applications. Table ~\ref{Parametros} shows the frequency bands of the seven channels, as well as the maximum transmission power permitted in each one.

The IEEE 802.11p standard defines four classes of devices, namely: Classes A, B, C, and D. These classes are characterized by a maximum transmission power, as shown in Table~\ref{Densidad} \cite{DSRC}.  The standard considers a spectral mask for each class of devices to reduce the radiated out-of-band power. Figure \ref{Masks} shows the shape and peak value of the spectral masks of the seven channels in the IEEE 802.11p standard  \cite{DSRC}. As can be seen in the figure, the power spectral density (PSD) is small, albeit not negligible, outside the limits of each channel's frequency band. This creates a problem of ACI. 

\subsection{QoS Evaluation}

\begin{figure}[htb]
	\centering
	\includegraphics[width=1\linewidth]{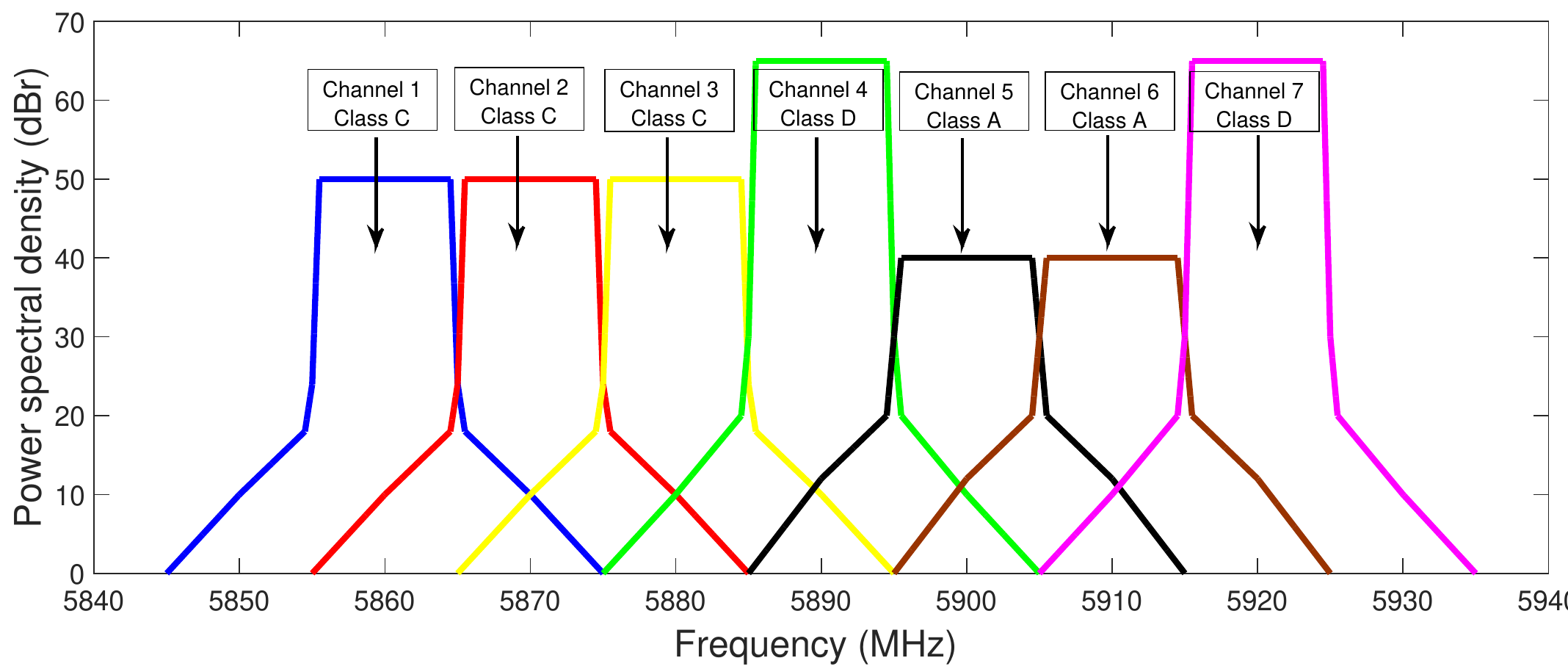}
	\vspace{-0.6cm}
	\caption{Spectral masks for 10 MHz channels in the IEEE 802.11p standard \cite{DSRC}.}
	\label{Masks}
\end{figure}

In this work, the QoS is evaluated through the SINR, as proposed in \cite{PowerControlForBroadcastV2V}. The SINR  {quantification} for the uplink channel from the $ i $-th OBU to the $ l $-th RSU at the time instant $k$ is given as:
	\begin{equation} 
	\label{eq1}
	\hat{\gamma}_{l,i}[k] 
	=\dfrac{W}{r_{l,i}}
	\frac{p_{l,i}[k] | h_{l,i}[k]| ^{2} }{  I_{l,i} [k] +   \sum_{m=1,m\neq l}^{M}p_{m,i}[k] |h_{m,i}[k]|^{2}  +\sigma^{2}_{l,i}} ~,
	\end{equation}
for all $ i \in [1,U]$, and $l \in [1,M]$, where $W$, $r_{l,i} $, $p_{l,i}[k]$ and $ h_{l,i}[k] $ denote the channel bandwidth, the data transmission rate, the transmission power and the complex-valued time-varying  gain of the mobile radio propagation channel associated to the link between the $ i $-th OBU and the $ l $-th RSU, respectively. The noise variance is denoted by $\sigma_{l,i}^2$, and $ \sum_{m=1,m\neq l}^{M}p_{m,i}[k] |h_{m,i}[k]|^{2} $  is the RSU to RSU interference. $I_{l,i} [k]$ represents the $i$-th element of the interference vector $\boldsymbol{I}_l [k]$ of the $l$-th RSU. This vector is defined as
	\begin{equation} \label{interf}
	\boldsymbol{I}_l [k]=\left[\begin{matrix}
	c_{l,2} |h_{l,2}[k]|^{2}p_{l,2}[k]  \\
	c_{l,1}|h_{l,1}[k]|^{2}p_{l,1}[k]+ c_{l,3}|h_{l,3}[k]|^{2}p_{l,3}[k]  \\
	c_{l,2}|h_{l,2}[k]|^{2}p_{l,2}[k]+ c_{l,4}|h_{l,4}[k]|^{2}p_{l,4}[k]   \\
	c_{l,3}|h_{l,3}[k]|^{2}p_{l,3}[k]+ c_{l,5}|h_{l,5}[k]|^{2}p_{l,5}[k]   \\
	c_{l,4}|h_{l,4}[k]|^{2}p_{l,4}[k]+ c_{l,6}|h_{l,6}[k]|^{2}p_{l,6}[k]   \\
	c_{l,5}|h_{l,5}[k]|^{2}p_{l,5}[k]+  c_{l,7}|h_{l,7}[k]|^{2}p_{l,7}[k]  \\
	c_{l,6}|h_{l,6}[k]|^{2}p_{l,6}[k]   \\
	\end{matrix}\right]~,
	\end{equation}
with $c_{l,1}=c_{l,2}=c_{l,3}=2.847 \times 10^{-4}$, $c_{l,4}=1.830 \times 10^{-5}$, $c_{l,5}= 6.081 \times 10^{-3}$, $c_{l,6}=6.050 \times 10^{-3}$ and $c_{l,7}=1.821 \times 10^{-5}$.  {These interference parameters $\{ c_{l,1}, \ldots, c_{l,7}\}$ were computed by the intersection area of each spectral mask with respect to the 10 MHz bandwidth of adjacent channels in Fig. \ref{Masks}.}
	
\begin{table}[t] 
	\caption{Assignment of channels in the 5.9 GHz band for DSRC systems  \cite{DSRC}.} 	\label{Parametros}
	\centering	
	\vspace{-0.3cm}
	{{
			\begin{tabular}{|c|c|c|c|c|c|c|c|c|}
				\hline \hline
				\textbf{Channel } &\textbf{Type of} &\textbf{Transmission } &\textbf{Frequency }\\ 
				\textbf{number} &\textbf{channel } &\textbf{power level} &\textbf{range } \\
				 & &  {(dBm)} &  {(GHz)} \\\hline 
				172 (Ch1) & Service &  33  & 5.855-5.865   \\\hline 
				174 (Ch2) & Service &  33 & 5.865-5.875 \\\hline
				176 (Ch3) & Service &  33 & 5.875-5.885  \\\hline
				178 (Ch4) & Control &  44.8 & 5.885-5.895  \\\hline
				180 (Ch5) & Service &  23 & 5.895-5.905  \\\hline
				182 (Ch6) & Service &  23 & 5.905-5.915  \\\hline
				184 (Ch7) & Service &  40 & 5.915-5.925 \\\hline
				\hline 	
	\end{tabular}}}
\end{table}

\begin{table*}[htb] 
	\caption{Power spectral density limits for 10 MHz channels in the IEEE 802.11p standard \cite{DSRC}, where the standard defines four device classes: A, B, C, and D (maximum power level allowed for each channel during transmission).} \label{Densidad}
	\centering	
	{{\small
			\begin{tabular}{|l|l|l|l|l|l|}
				\hline \hline
				& \multicolumn{5}{c|}{Upper bound on the transmission power level} \\
				Power& \multicolumn{5}{c|}{ in relative decibels (dBr). } \\\cline{2-6}
				class & $\pm$ 4.5 MHz &  $\pm$ 5.0 MHz & $\pm$ 5.5 MHz &  $\pm$ 10 MHz &  $\pm$ 15 MHz\\ 
				& offset & offset & offset  &offset &offset\\\cline{1-6}
				A & 0 & -10 & -20 & -28&-40\\ \cline{1-6}
				B & 0 & -16 & -20 & -28&-40\\ \cline{1-6}
				C & 0 & -26 & -32 & -40&-50\\ \cline{1-6}
				D & 0 & -35 & -45 & -55&-65\\ \cline{1-6}
	\end{tabular}}}
\end{table*}

\subsection{ {V2I Propagation Channel Model}}

The mobile radio propagation channel for the link between the $ i $-th OBU and the $ l $-th RSU is modeled by
\begin{equation}
	h_{l,i}[k]=g_{l,i}[k] \lambda_{l,i}[k] \left(\dfrac{0.1}{d_{l,i}[k]}\right)^{ {\epsilon}}~,
	\label{modelo}
	\end{equation}
for all $i \in [1,U]$, and $l \in [1,M]$, where $g_{l,i}[k]$ and $\lambda_{l,i}[k]$ are discrete-time complex-valued stochastic processes characterizing the effects of fast fading due to multipath propagation, and shadowing by the presence of objects in the coverage area, respectively  {\cite{Carlos2018}}. The third term at the right-hand side of the previous equation stands for the propagation path loss, where $d_{l,i}[k]$ denotes the distance from the $ i $-th OBU to the $ l $-th RSU, and  {$\epsilon$ is the path-loss exponent}.  {For simplicity, the distances between OBUs and RSUs are computed by a two-dimensional approximation, where the heights of the objects are neglected. Nonetheless, a three-dimensional perspective could be also used just by adjusting the computation of $d_{l,i}[k]$. Following \cite{Power_Control_in_D2D-based_Vehicular_2015}, we set the path-loss exponent $\epsilon$ equal to 3.}

We model  {the fast fading component} $g_{l,i}[k]$ by the superposition of $N_c$ plane waves as follows
\begin{equation}
	g_{l,i}[k] = \sum_{n=1}^{N_{c}}c_{n} \exp \{\iota({2\pi f_{n}k+\theta_{n}})\} \qquad \iota=\sqrt{-1}~,
	\end{equation}
where $ c_{n} $, $f_{n} $, and $\theta_{n}$ are the amplitude, the Doppler shift, and the phase of the $n$-th received plane wave \cite{gutierrez2010design}. The phase terms $\theta_{n}$ are characterized by independent and identically distributed (i.i.d.) random variables, each having an uniform distribution over $[-\pi, \pi)$.  In turn, the Doppler frequencies $f_{n}$ are defined as:
\begin{equation}
	f_{n} \overset{\Delta}{=} f_{max} \cos (\alpha_{n}) \qquad \forall n \in [1,N_{c}]~,
	\end{equation} 
where $f_{max}= \frac{v}{\lambda}$ is the maximum Doppler shift caused by the movement of the $i$-th OBU, $v$ is the speed of the OBU, $\lambda$ is the wavelength, and $\alpha_{n}$ stands for the angle of arrival (AOA) of the $n$-th incident wave. We assume that the AOAs  {$\{\alpha_ {n}\}_{n=1}^{N_c}$} are i.i.d. random variables uniformly distributed over $[0, \pi)$. In our system model, the carrier frequency $f_c$ takes different values given by the central frequency of the DSRC bands described in Table ~\ref{Parametros}. Meanwhile, the shadowing effects are characterized according to 
\begin{equation}
	\lambda_{l,i}[k]=10^{(\sigma_{L}\sum_{n=1}^{N_{c}}c_{n}\cos(2\pi f_{n}k+\theta_{n})+m_{L})/20}~,
\end{equation}
where $ \sigma_ {L} $ and $ m_{L} $ represent the standard deviation and average value of the shadow fading process, respectively \cite{patzold2011mobile}.

\subsection{ {SINR Estimation}}

The QoS of the communication link between the $ i $-th OBU and the $ l $-th RSU is guaranteed at the time instant $k$  if $ \gamma_{l,i}[k] \geq \gamma_{l,i}^{obj} $, where $\gamma_{l,i}^{obj} $ is the objective SINR threshold that allows achieving the desired QoS. An important aspect of the SINR modeled by \eqref{eq1} is that at any given time instant, the objective SINRs $\{ \gamma_{l,i}^{obj} \}$ can be achieved by computing the transmission powers $\{p_{l,i}[k]\}$ that solve a linear system of  {$M \cdot U$} equations \cite{Unified}. This solution establishes a one-to-one relation among objective SINRs and OBU transmission powers, although it is centralized and requires real-time feedback to the OBUs.

The SINR  {measurement} in (\ref{eq1}) has some variability by the estimates of the channel and the  {interference}  factors, and to reduce this effect, we apply a filtering stage to the estimation ${\hat \gamma}_{l,i}[k]$ in \eqref{eq1}.  For this purpose, an $ \alpha- \beta- \gamma $ filter is used to compute $ \gamma_{l,i}[k]$ \cite{filter}, which is commonly employed in radar systems. The dynamic equations given for the $ i $-th OBU and $ l $-th RSU  are \cite{filter, Simon}:
\begin{align}
x_{l,i}[k+1]& =\gamma_{l,i} [k]+T_s\, v_{l,i,s}[k]+\frac{1}{2}T_s^{2}a_{l,i,s}[k]~, \label{filter1}\\
v_{l,i,p}[k+1] & =v_{l,i,s}[k]+T_s\, a_{l,i,s}[k]~,
\end{align}
where the states ($x_{l,i}, v_{l,i,p}$) are employed to compute the smoothed
signals $ \gamma_{l,i}[k] $, $ v_{l,i,s}[k] $  {and} $ a_{l,i,s}[k]$ as follows:
\begin{align}
\gamma_{l,i}[k] & =x_{l,i}[k]+\alpha (\hat{\gamma}_{l,i}[k]-x_{l,i}[k]), \label{gammaf}\\
v_{l,i,s}[k]  & =v_{l,i,p}[k]+\frac{\beta}{T_s} (\hat{\gamma}_{l,i} [k]-x_{l,i}[k])~,\\
a_{l,i,s}[k] & =a_{l,i,s}[k-1]+\frac{\gamma}{2T^{2}_s} (\hat{\gamma}_{l,i} [k]-x_{l,i}[k]) ~, \label{filter2}
\end{align}
$ T_s$  is the sampling interval for power allocation, and parameters $ \alpha $, $\beta$ and $\gamma $ influence the behavior of the filter in terms of stability and time response.

\section{Network Utility Optimization}

In this work, the concept of utility in the communication network focuses on energy efficiency, so it evaluates the total number of bits transmitted successfully per Joule of energy consumed \cite{pricing, Poor}. On one hand, a high level of SINR between $ i $-th OBU and $ l $-th RSU will result in a lower bit error rate (BER) and therefore on a higher transmission rate. On the other hand, reaching a high level of SINR requires that the OBUs transmit at a higher power level, resulting in greater interference between adjacent channels, and in larger interference from RSU to RSU.

\subsection{ {Utility Function}}

This trade-off can be quantified by a utility function for the $ i $-th OBU in $ l $-th RSU at $k$-time instant as \cite{Poor, PreEqualization}:
\begin{equation} \label{uil}
	u_{l,i}[k]= \dfrac{T_{l,i}[k]}{p_{l,i}[k]} \quad \frac{bits}{joule}~ \quad \forall i \in [1,U]~, \; l \in [1,M]~,
\end{equation}
where $T_{l,i}[k]$ defines the corresponding throughput for the active OBU. Moreover, this utility function can be expressed as: 
\begin{equation}
	u_{l,i}[k]= \underbrace{\dfrac{L r_{l,i}f(\gamma_{l,i}[k])}{N}}_{T_{l,i}[k]}  \dfrac{1}{p_{l,i}[k]}= \underbrace{\dfrac{L r_{l,i}}{N}}_{w_{l,i}}  \dfrac{f(\gamma_{l,i}[k])}{p_{l,i}[k]} \quad \dfrac{bits}{joule}~,
	\label{UtiDef}
\end{equation}
where $L$ is the number of information bits per orthogonal frequency division multiplexing (OFDM) symbol, and  $N$ the total number of bits in each OFDM symbol ($ N>L $); $r_{l,i}$ is the data rate, and $f(\gamma_{l,i}[k]) $ is an efficiency function related to the reception rate of OFDM symbols. In general, the efficiency function $ f( \gamma_{l,i}[k])$ has the following properties: 
\begin{itemize}
		\item $f(0) \to 0$ and $f(\infty) \to 1$,
		\item $f'(\gamma_{l,i}[k]) > 0 $ $\forall \gamma_{l,i}[k] > 0$ with $f'(0) \to 0 $ and $f'(\infty) \to 0$.
\end{itemize}
One type of efficiency function is given in \cite{PreEqualization}:
\begin{equation}
	f(\gamma_{l,i}[k])=(1-\exp^{-\gamma_{l,i}[k]})^{N}~,
	\end{equation}
therefore, the overall network utility can be written as:
\begin{align}
\mathcal{U}(\mathbf{p}_{1}[k],\cdots,\mathbf{p}_{M}[k]) & =  \sum_{l=1}^{M} \sum_{i=1}^{U} u_{l,i}[k] \nonumber \\
& = \sum_{l=1}^{M} \sum_{i=1}^{U} w_{l,i}\dfrac{f(\gamma_{l,i}[k])}{p_{l,i}[k]}, \label{Uopt}
\end{align}
where
\begin{equation}
	\mathbf{p}_{l}[k]=[p_{l,1}[k],\ldots,p_{l,U}[k]]^{ \top}  \quad \forall l \in [1,M]~.
\end{equation}

\subsection{ {Utility Optimization}}

 The proposed optimization problem determines the values of the vectors $\{\mathbf{p}_l[k]\}_{l=1}^M$ at $k$-time instant that maximize the network utility under a specific QoS interval. Mathematically, the optimization problem can be described as follows:
\begin{equation} \label{eq8}
\max_{\{\mathbf{p}_{1}[k], \ldots, \mathbf{p}_{M}[k] \}}  { \mathcal{U}(\mathbf{p}_{1}[k], \ldots, \mathbf{p}_{M}[k]) }~, 
\end{equation}
\begin{equation}  \nonumber
\quad \textrm{s.t.} \quad  \gamma_{min}  \leq  \gamma_{l,i}[k]  \leq  \gamma_{max} \; \; \forall i \in [1,U], \; l \in [1,M]~,
\end{equation}
where $\gamma_{min}$ and $\gamma_{max}$ are the lower and upper bound, respectively, in the SINR which are related to  QoS.  {The stationary conditions for \eqref{Uopt} are analytically derived in the Appendix}.

\begin{figure}[htb]
	\centering
	\includegraphics[width=1\linewidth]{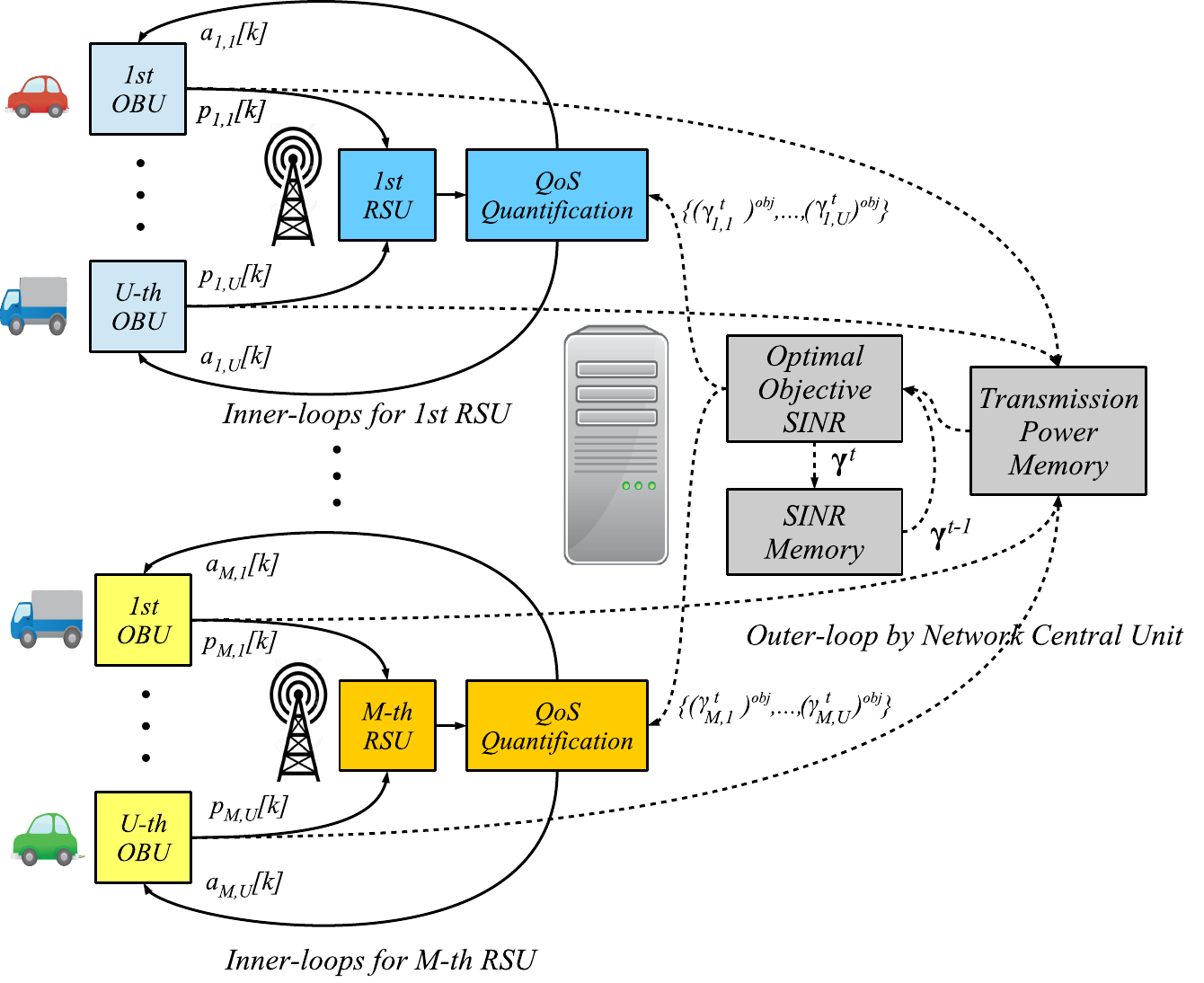}
	\caption{Proposed  {two-loop feedback structure for network utility maximization, where the outer-loop is implemented in a network central unit, and the inner-loops in the OBUs-RSUs}.}
	\label{DiagUtility}
\end{figure}

 {A distributed solution to \eqref{eq8} is deduced by the optimality equations in the Appendix, and by the injective relation between objective SINR per OBU and its transmission power in a communication system with interference \cite{Unified}. Hence we propose to perform the optimization process by an iterative and distributed scheme which uses two adjustment loops at different time scales (as shown in Fig. \ref{DiagUtility}):}
\begin{itemize}
\item \textbf{Outer-loop}: we introduce a new time index $t$ (slow-time scale) for this loop, with respect to the original  {$k$-time instant}, i.e. $t=\lfloor k/Q \rfloor$ where $\lfloor \cdot \rfloor$ denotes the floor function, and $Q>0$ the number of time samples in each iteration of the outer-loop. The optimal SINR per OBU is computed by the following equation:
 \begin{equation}
	f(\gamma_{l,i}^{t})\gamma_{l,i}^{t}-f(\gamma_{l,i}^{t})= \hat{M}_{l,i}(\boldsymbol\gamma^{t-1},\textbf{p}^{t-1})~,
	\label{sinrobj}
\end{equation}
where the SINR values $\boldsymbol{\gamma}^{t-1}$ and estimated transmission power $\mathbf{p}^{t-1}$ in the previous iteration are used to have a constant value in the right-hand side of \eqref{sinrobj}, and  {$\hat{M}_{l,i}(\cdot)$ in \eqref{Mli} is derived in the Appendix}. With this simplification, the equations in \eqref{sinrobj} are decoupled and their solutions can be easily reached by a numerical algorithm  {by a network central unit}. To estimate the transmission power in the $t-1$ iteration $\mathbf {p}^{t-1}$, an average over the time index $ t-1 $ is considered:
\begin{equation}
	\mathbf{p}^{t-1}=\dfrac{1}{Q} \left[ \begin{array}{c}
	\sum_{k=k_o}^{Q+k_o}p_{1,1}[k]  \\ 
	\vdots \\
	\sum_{k=k_o}^{Q+k_o}p_{M,U}[k]
	\end{array} \right]~, 
	\label{VectorPromedio}
	\end{equation}
where $k_o=(t-1) \cdot Q$ represents the sample time at which the previous $t-1$ update window started. Once (\ref{sinrobj}) is solved for each OBU, the target SINR value is found $(\gamma_{l,i}^t)^{obj}$ by considering the lower and upper bounds in \eqref{eq8}:
\begin{equation}
	(\gamma_{l,i}^{t})^{obj}=\max (\gamma_{min}, \min [\gamma_{l,i}^{t}, \gamma_{max} ])~.
	\label{SINRoptima}
\end{equation}
This objective SINR $(\gamma_ {l, i} ^ {t}) ^ {obj}$ defines the required QoS in the V2I communication link to maximize the network utility.

\item \textbf{Inner-loops}: the internal control loops are proposed to assign the transmission power $p_ {l, j} [k] $ of each OBU to reach the objective SINR $ (\gamma_{l, i}^{t})^{obj}$ based on a feedback structure at $k$-time instant.  Thus, the inner-loops are responsible to guarantee the desired QoS in the V2I communication link despite channel variations, and network latency.  {The state of the art in distributed power control algorithms for wireless networks is vast \cite{Chiang, DistSNR, Unified, LATAM2013}. In \cite{LATAM2013}, the robustness to round-trip delay uncertainty, and reference tracking performance was studied for seven power control schemes: (i) Fixed-step, (ii) Foschini-Miljanic, (iii) proportional-integral-derivative control, (iv) $H_\infty$ robust control, (v) Robust Smith predictor, (vi) Variable structure control, and (vii) linear-quadratic-gaussian (LQG) control. This study showed that the best compromise is achieved by the LQG control in \cite{Unified}. On the other hand, the results in \cite{DistSNR} show that this power control scheme presents the optimal solution to improve robustness to quantization and measurement noise in wireless networks. These previous studies highlight that the LQG control is the best choice for the inner-loop structure.} In the LQG control scheme, the $l$-th RSU estimates the following QoS error signal for the $i$-th OBU:
\begin{equation}
	e_{l,i}[k]=\left[\frac{(\gamma_{l,i}^{t})^{obj}}{\gamma_{l,i}[k]} - 1 \right] p_{l,i}[k]~,
	\label{ErrorPorcentual}
\end{equation} 
and this information is sent to the OBU to update its transmission power level.  In addition, there is a parameter denoted as $ \Omega \in (0,1) $ that balances the control effort against the tracking convergence speed  \cite{Unified, LATAM2013}.  In this way, the LQG power allocation scheme for $i$-th OBU at $l$-th RSU is expressed by:
\begin{equation} \label{LQG}
	p_{l,i}[k+1]=(1-\Omega)p_{l,i}[k] + \Omega p_{l,i}[k-n_{RT}] - \Omega a_{l,i}[k],
\end{equation}
where $a_{l,i}[k] $ denotes the QoS error signal received at the $ i $-th OBU from the $l$-th RSU. This error signal is affected by $n_{RT}$ time delays linked to the latency in the power update and QoS quantification, its expression is given by:
\begin{equation}
	a_{l,i}[k]=e_{l,i}[k-n_{RT}].
	\label{error}
\end{equation}
Hence, the main advantage of the LQG power allocation in \eqref{LQG} is that it explicitly includes the effect of the time delays / network latency $n_ {RT}$ in its structure.
\end{itemize}

\subsection{ {Computation Complexity}}

The complexity of the proposed distributed mechanism for network utility maximization in Fig. \ref{DiagUtility} can be analyzed with respect the objectives of the two-loop structure: (i) the computation of the optimal SINRs in \eqref{sinrobj} by the outer-loop, and (ii) the power update mechanisms in \eqref{LQG} by the inner-loops. In addition, the overall complexity has to consider that the two-loop mechanism is executed at two time scales characterized by indexes $t$ and $k$. Hence, between each iteration of the outer-loop, the $l$-th RSU requires a continuous information sharing of the resulted transmission powers $\{\mathbf{p}_l[k_o],\ldots,\mathbf{p}_l[Q+k_o]\}$ in the $U$ OBUs for $Q$ samples to a central unit in the network to compute the average in \eqref{VectorPromedio}. Next, at $t$-iteration of the outer-loop, $M \cdot U$ equations in \eqref{SINRoptima} are solved for the optimal SINRs by the central network unit.  However, this task is not time consuming since the $M \cdot U$ equations are decoupled among them, and they can be efficiently solved by root-finding numerical methods, since the left-hand side in \eqref{sinrobj} is a smooth function. These  solutions $\{ \gamma_{1,1}^t, \ldots, \gamma_{M,U}^t \}$ are then limited to the SINR feasible interval in \eqref{SINRoptima}. From the network central unit, the objective SINRs $\{ (\gamma_{1,1}^t)^{obj}, \ldots, (\gamma_{M,U}^t)^{obj}\}$ are fedback to the $M$ RSUs in the network, and also saved in memory for the next iteration of the outer-loop.

Meanwhile, there are $M \cdot U$ inner-loops that are implemented in the network to adjust the transmission power according to \eqref{LQG} at each $k$-time instant. Each RSU computes $U$ error signals in \eqref{ErrorPorcentual} by using the SINR estimation $\gamma_{l,i}[k]$ in \eqref{gammaf} coming from the $ \alpha- \beta- \gamma $ filter, the objective SINR $(\gamma_{l,i}^t)^{obj}$, and the transmission power $p_{l,i}[k]$  (QoS Quantification blocks in Fig. \ref{DiagUtility}). As a result, each RSU has to run in parallel $U$ filters in \eqref{filter1}-\eqref{filter2} to estimate the SINRs in each OBU.  Meanwhile, the raw SINR measurement $\hat{\gamma}_{l,i}[k]$ in \eqref{eq1} could be computed by the RSUs from either a direct substitution of the channel gains, transmission powers, noise variance, data transmission rate and channel bandwidth, or by pilot-based time or frequency  domain estimation \cite{Ren,Das}. Finally, the OBUs implement the power adjustment algorithm in \eqref{LQG} that involves $n_{RT}$ memory units, three scaling operations and two summations. In the overall, the complexity of the proposed utility maximization methodology in Fig. \ref{DiagUtility} is distributed among the network central unit, $M$ RSUs and $M \cdot U$ OBUs.

\section{Simulation Evaluation}

The proposed scheme to maximize the network utility in Fig. \ref{DiagUtility} was evaluated through numerical simulation in the uplink of a VCN, where three omnidirectional RSUs of 1 km radius in the network are considered, i.e. $M = 3$,  {with seven OBU's linked to each RSU, i.e. $U=7$. Each part of our simulation platform was initially validated as an isolated system. This validation process included the individual testing of our small-scale simulator, large-scale fading simulators, mobility profile simulator, and power control algorithms \cite{Carlos2018,gutierrez2010design,patzold2011mobile,Unified, LATAM2013}. Then, we conducted several trials to make sure that the numerical results produced by the complete simulation platform were consistent}. The parameters of the simulation are illustrated in Table \ref{PaSiReVe}, where the numerical implementation was carried out in Matlab. 

 {The simulation platform is initialized by considering a distribution of the OBUs as shown in Fig. \ref{Modelo}. In the mobility model, we are considering a straight road scenario that is typically found in urban, semi-urban and rural environments, where the OBUs have a constant velocity. In the testing conditions, the minimum safety distance among neighboring OBUs is 10 m.} During the first 50 iterations of the  {inner-loops}, the initial power for all OBUs is set at 1 pW, and the objective SINR is constant for them at 5 dB, which results in a data transmission rate of 3 Mbps \cite{OptimalData}.  Furthermore, during the evaluation, the time-delay related to QoS quantification and network latency was varied between 0 and 10 samples with an uniform distribution, where the time-delay is updated every 20 samples (i.e. every second). This condition is adopted since  {vehicular} communications in practice face a maximum latency of 500 ms \cite{Vehicular_Networking}, which corresponds to 10 samples in this work. With this consideration, in the LQG algorithm in (\ref {LQG}), an estimated value of $n_{RT} = 5$ is considered.  {In the overall network,} the VCN setup considers 21 active OBUs and 3 RSUs with interference between adjacent channels, and  interference from RSU to RSU. Three different propagation scenarios are  {validated} for the OBUs mobility:
\begin{itemize}
	\item  \textbf{Scenario A}:  some OBUs approach and others drive away from the RSU, as shown in Fig. \ref{Modelo}.
	\item  \textbf{Scenario B}:  all the OBUs approach simultaneously to the RSU.
	\item  \textbf{Scenario C}:  all the OBUs move away from the RSU. 
\end{itemize}
To compare the utility improvements by the time-varying objective SINRs in \eqref{SINRoptima}, the same three scenarios are evaluated with  {fixed objective SINRs in the inner-loops of 5, 7, 9 and 11 dB for all OBUs in the network}.

\begin{table}[htb] 
	\caption{Parameters of the VCN during the  {simulation} evaluation.} \label{PaSiReVe}
	\centering	
	{{\small
			\begin{tabular}{|l|l|l|l|l|l|l|l|l|}
				\hline \hline
				\textbf{Parameter} &\textbf{Variable } &\textbf{Value }\\\hline 
				Total number of OBUs & $ U $ & 21 \\\hline 
				Total number of RSUs & $M$ & 3 \\\hline 
				Noise power & $ \sigma^{2} $ & -90 dBm  \\\hline 
				Number of bits per OFDM symbol & $ N$ & 64  \\\hline
				Number of information bits per per  & $ L$ & 48 \\
				OFDM symbol & & \\\hline
				Bandwidth per channel & $ W$ & 10 MHz  \\\hline
				Data rate & $ r_{l,i}$ & 3 Mbps  \\\hline
				Control Gain LQG & $\Omega$  & 0.10 \\\hline
				Frequency of power control update & $f_{1}$ & 20 Hz \\\hline 	
				Frequency of objective SINR update & $f_{2}$ & $ 2/5 $ Hz \\\hline
				 {Update period of outer-loop} &  {$Q$} &  {50} \\ \hline 
				Cell coverage radius & $R$ & 1 km \\\hline
				Duration of the simulation& $t$ & 25 s\\\hline 
				Maximum transmission power &$p_{max}$ & 30.2 W \\\hline
				Minimum transmission power &$p_{min}$ & 1 $p$W \\\hline
				Minimum safety distance & $d_{min}$ & 10 m \\\hline
				Distance from highway to RSU & $D$ & 150 m \\\hline
				Loss-exponent &$\epsilon$ & 3 \\\hline
				Maximum delay &$n_{RT}^{max}$  & 10 samples\\\hline
				Minimum delay &$n_{RT}^{min}$  & 0 samples\\\hline
				Estimated delay for the LQG control & $n_{RT}$  & 5 samples\\\hline
				Maximum latency& $L_{max}$ & 500 ms\\\hline
				Gains of smoothing filter &$\alpha$  & 0.4 \\
				 &$\beta$ & 0.001\\
				&$\gamma$  & $2 \times 10^{-5}$\\\hline
				Standard deviation of the shadow  & $ \sigma_{L} $ & 6 dB \\
				fading &  & \\\hline
				Mean of the shadow fading & $ m_{L} $  & 0\\\hline
				\hline 		
	\end{tabular}}}
\end{table}

\subsection{Scenario A}

\begin{figure}
	\centering
	\includegraphics[width=1\linewidth]{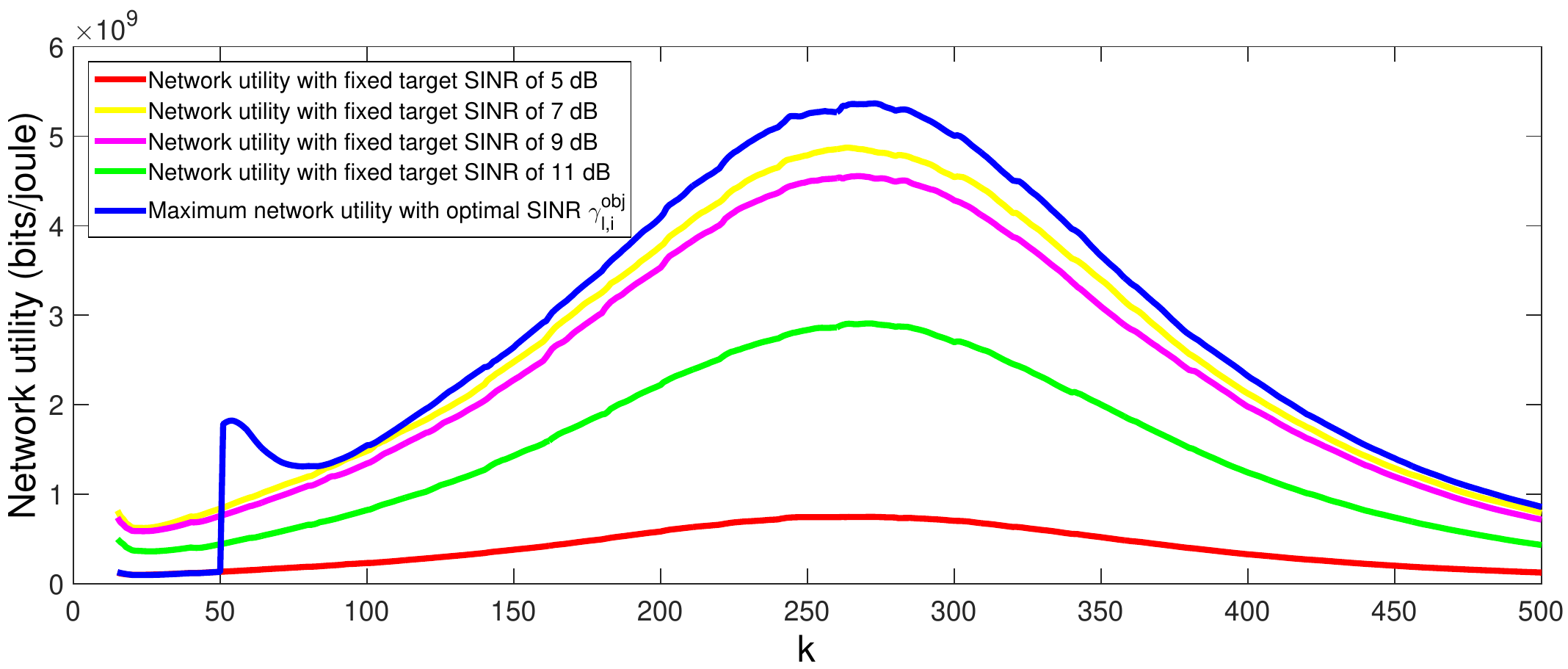}
	\caption{ {\textbf{Scenario A}: Monte Carlo evaluation of the instantaneous network utility in \eqref{Uopt}, with 100 closed-loop realizations of 500 samples total duration, and OBUs speed of 72 km/h} (the OBUs approach and move away from the RSUs).}
	\label{utilidadredA72}
\end{figure}

\begin{figure}
	\centering
	\includegraphics[width=1\linewidth]{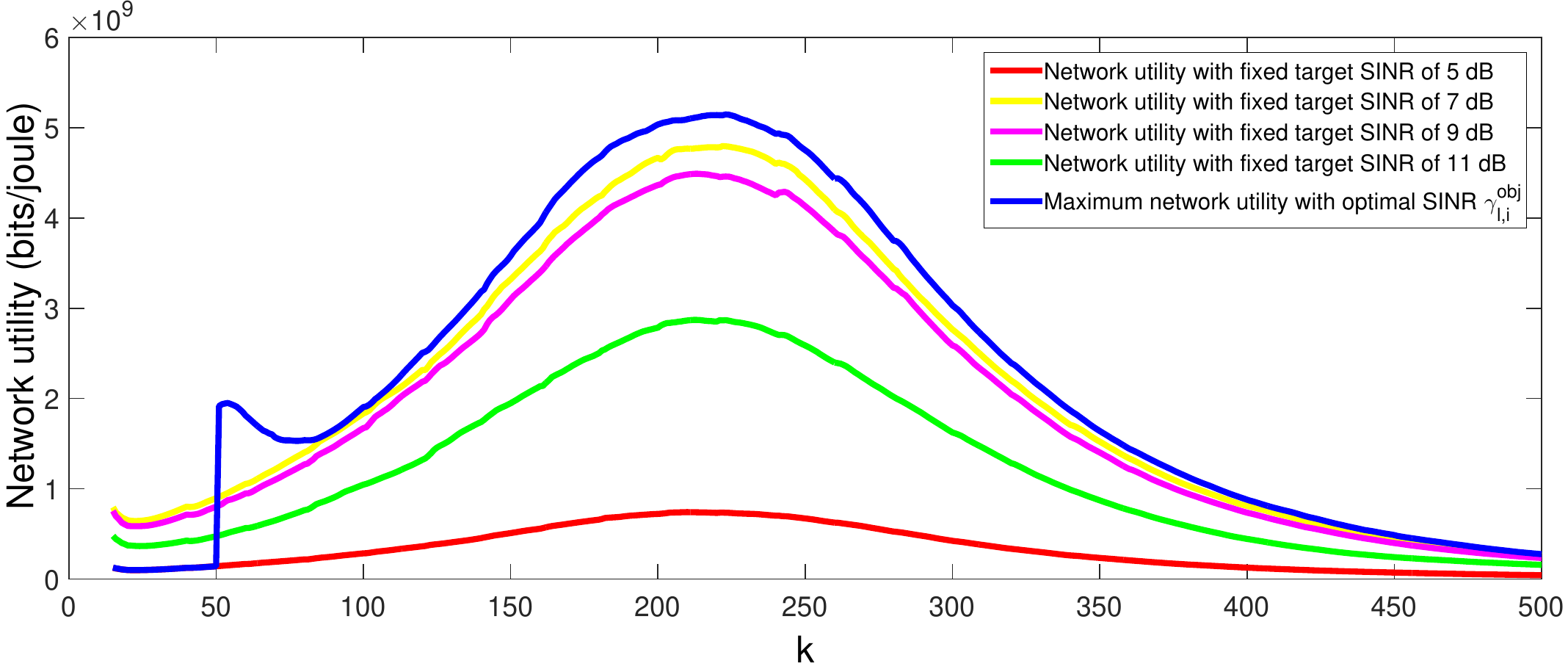}
	\caption{ {\textbf{Scenario A}: Monte Carlo evaluation of the instantaneous network utility in \eqref{Uopt}, with 100 closed-loop realizations of 500 samples total duration, and OBUs speed of 90 km/h} (the OBUs approach and move away from the RSUs).}
	\label{utilidadredA90}
\end{figure}

\begin{figure}
	\centering
	\includegraphics[width=1\linewidth]{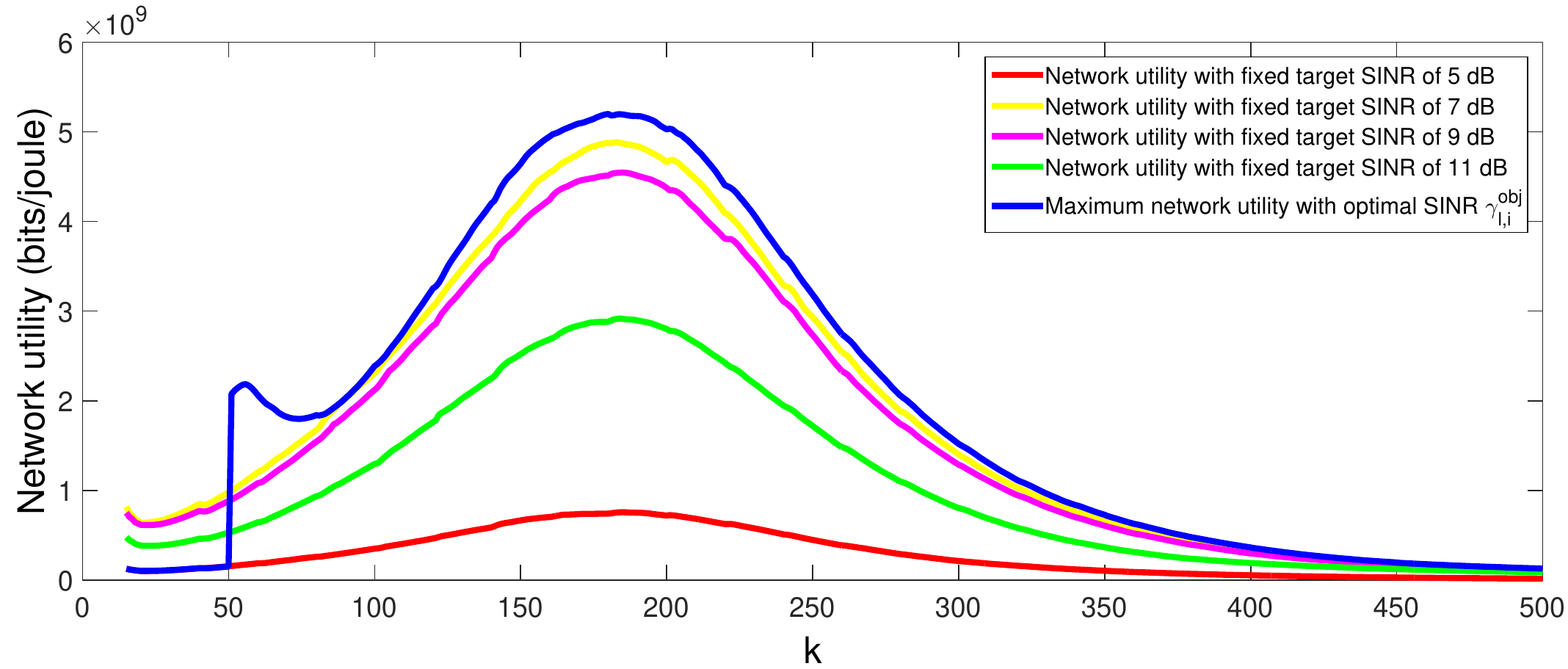}
	\caption{ {\textbf{Scenario A}: Monte Carlo evaluation of the instantaneous network utility in \eqref{Uopt}, with 100 closed-loop realizations of 500 samples total duration, and OBUs speed of 108 km/h} (the OBUs approach and move away from the RSU).}
	\label{utilidadredA108}
\end{figure}

In this scenario, for the described mobility profile, we consider three constant propagation speeds for the OBUs: a) 72 km/h, b) 90 km/h, and c) 108 km/h. In our simulations, the large-scale and small-scale fadings in \eqref{modelo} are updated at each  {$k$-time instant}. For comparison purposes, a Monte Carlo evaluation is presented with 100 closed-loop realizations of 500 samples each one, where the network utility in \eqref{Uopt} is computed every time sample. The results of this evaluation are presented in Figs. \ref{utilidadredA72}, \ref{utilidadredA90} and \ref{utilidadredA108}. As expected, the maximum network utility is accomplished by our proposal in Fig. \ref{DiagUtility} with the optimum SINR in (\ref {SINRoptima}), and despite the time-varying mobility profile with different propagation speeds. Note that during the first 50 samples, the objective SINR is fixed at 5 dB, and next, the optimal time-varying SINR is updated every 50 samples (i.e. $Q=50$). For this reason, after the first 50 samples, a large transient is observed in the network utility for the optimal strategy in Figs. \ref{utilidadredA72}, \ref{utilidadredA90} and \ref{utilidadredA108},  {compared to the fixed objective SINR conditions. In these three figures, we observe that for all objective SINR conditions, the network utility achieves a peak during the 500 samples of the simulation, which corresponds to the time of minimum distance of the OBUs to the RSU.  As a result, these peak utility conditions are achieved at different time instants which are modified by the propagation speeds, i.e. as the speed is larger, the peak utility is reached earlier.}  


\begin{figure}
	\centering
	\includegraphics[scale=0.22]{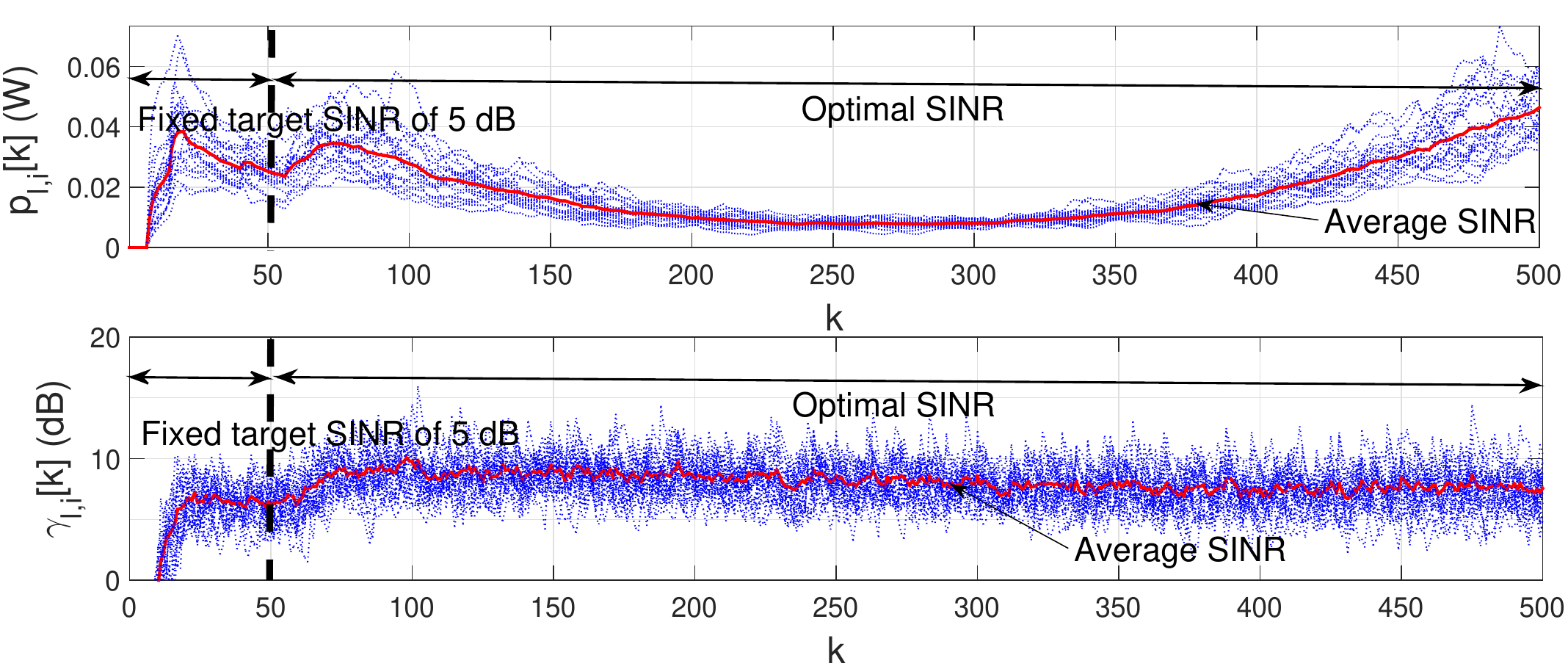}
	\vspace{-0.35cm}
	\caption{ {\textbf{Scenario A}: Sample of the LQG power allocation in the inner-loops with OBUs} speed of 72 km/ h (the OBUs approach and move away from the RSU):  (bottom) instantaneous SINR, and (top) transmission power level.}
	\label{powerSINR72A}
\end{figure}

\begin{figure}
	\centering
	\includegraphics[scale=0.22]{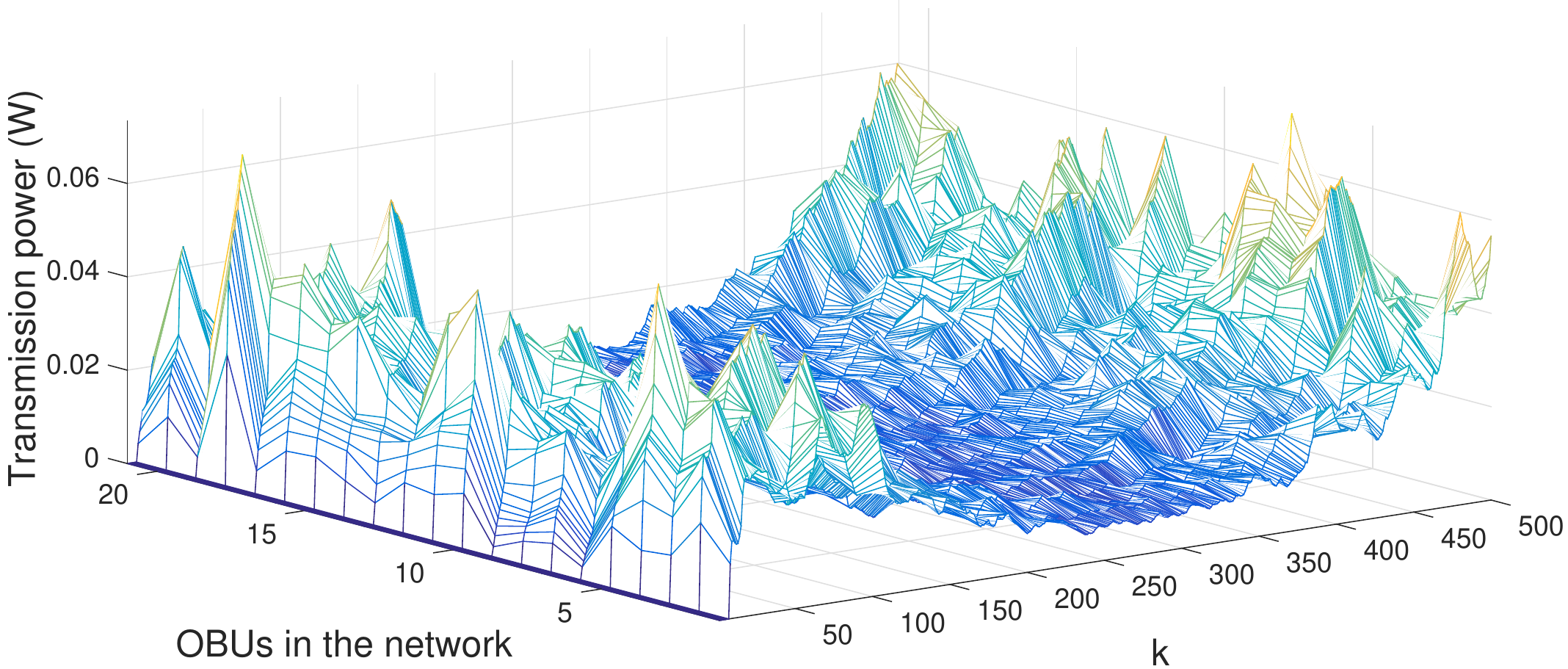}
	\vspace{-0.4cm}
	\caption{ {\textbf{Scenario A}: Sample of the transmission power per OBU in the inner-loops with} speed of 72 km/h (the OBUs approach and move away from the RSU).}
	\label{Power72A}
\end{figure}

Figure \ref {powerSINR72A} illustrates one realization of the time responses for the LQG power allocation   {in \eqref{LQG}}   {for the inner-loops in Scenario A and OBUs} speed of 72 km/h. In fact, the corresponding results at 90 km/h and 108 km/h show the same time pattern, but are omitted for brevity. In Fig. \ref {powerSINR72A}, the top panel presents the required transmission power level for each OBU throughout the simulation time by the LQG control law, as well as, the average power level for all OBUs; meanwhile, the bottom panel shows the achieved SINR for each OBU, and also the average  {SINR} response. Hence, in the first 50 samples, the objective SINR is fixed at 5 dB for all OBUs and after this initialization period, each OBU has a different target for the rest of the simulation time to maximize the network utility  {in \eqref{Uopt}}. Note that the optimal objective SINRs are always lower to 10 dB, but as shown in Fig. \ref{utilidadredA72}, the obtained network utility is larger  {than for} the constant objective SINR conditions  {of 5, 7, 9 and 11 dB}. Due to the fast channel variations, the SINR produces small oscillations around its objective value. This effect can be visualized in the bottom panel of Fig. \ref {powerSINR72A}, where the time-varying channel gains produce perturbations in the  {SINR} tracking performance, but the LQG scheme is always able to compensate these variations. In addition, Fig. \ref{Power72A} illustrates the resulting power per OBU, where due to the time-varying mobility profiles, the instantaneous transmission power has a minimum value when the OBUs are close to the RSUs, and at the end of the simulation, this value is raised since the OBUs are moving away from them. In this plot, the dynamic property of the LQG power allocation scheme  {in \eqref{LQG}} compensates the fast and slow channel fadings, and the OBU mobility profiles to track the objective SINRs. 

\subsection{Scenario B}

\begin{figure}
	\centering
	\includegraphics[width=1\linewidth]{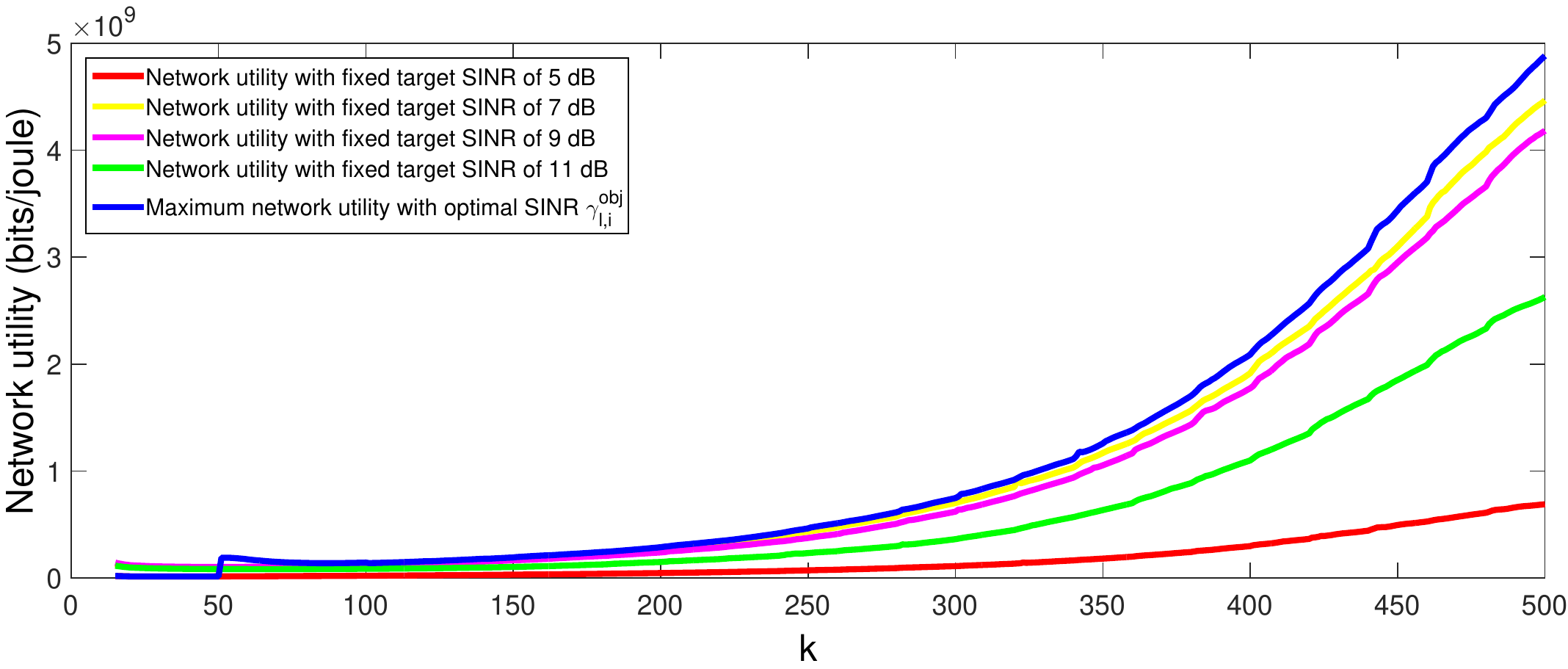}
	\caption{ {\textbf{Scenario B}: Monte Carlo evaluation of the instantaneous network utility in \eqref{Uopt}, with 100 closed-loop realization of 500 samples total duration, and OBUs speed of 72 km/h} (all the OBUs approach the RSU).}
	\label{utilidadredB}
\end{figure}

A new Monte Carlo evaluation is considered with also 100 closed-loop realizations and 500 samples in each one for a speed of 72 km/h. Once more, for brevity, the responses at constant speeds of 90 km/h and 108 km/h are omitted, but the performance is consistent with Scenario A.  The results in Fig. \ref{utilidadredB} show that our optimal scheme allows to maximize the network utility compared to fixed objective SINRs  {of 5, 7, 9 and 11 dB} in Scenario B. Our results illustrate that the network utility keep increasing during all the simulation time, since the OBUs are approaching to the RSUs at a constant speed, and consequently, the channel response is improving continuously.

\begin{figure}
	\centering
	\includegraphics[scale=0.22]{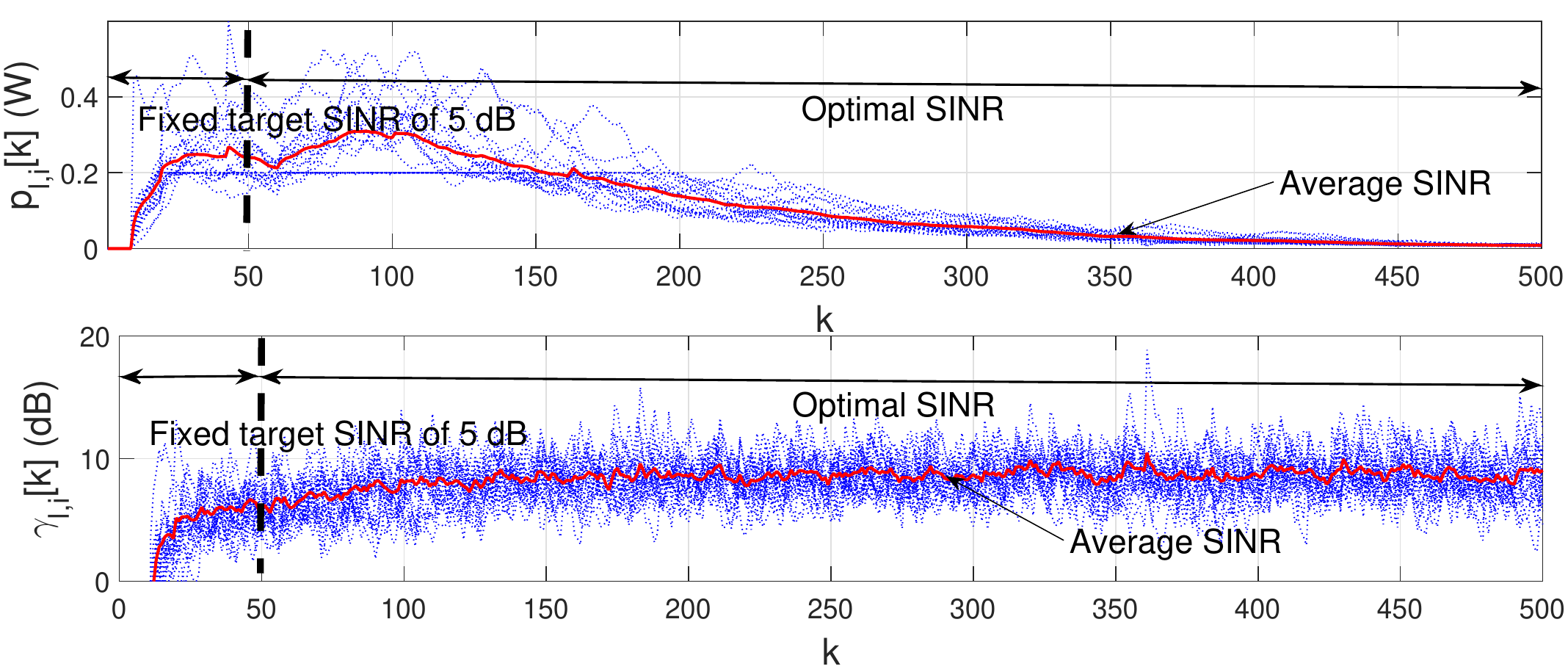}
	\vspace{-0.35cm}
	\caption{ {\textbf{Scenario B}: Sample of the LQG power allocation in the inner-loops with OBUs} speed of 72 km/ h (all the OBUs approach the RSU): (bottom) instantaneous SINR, and (top) transmission power level.}
	\label{powerSINR72B}
\end{figure}

\begin{figure}
	\centering
	\includegraphics[scale=0.22]{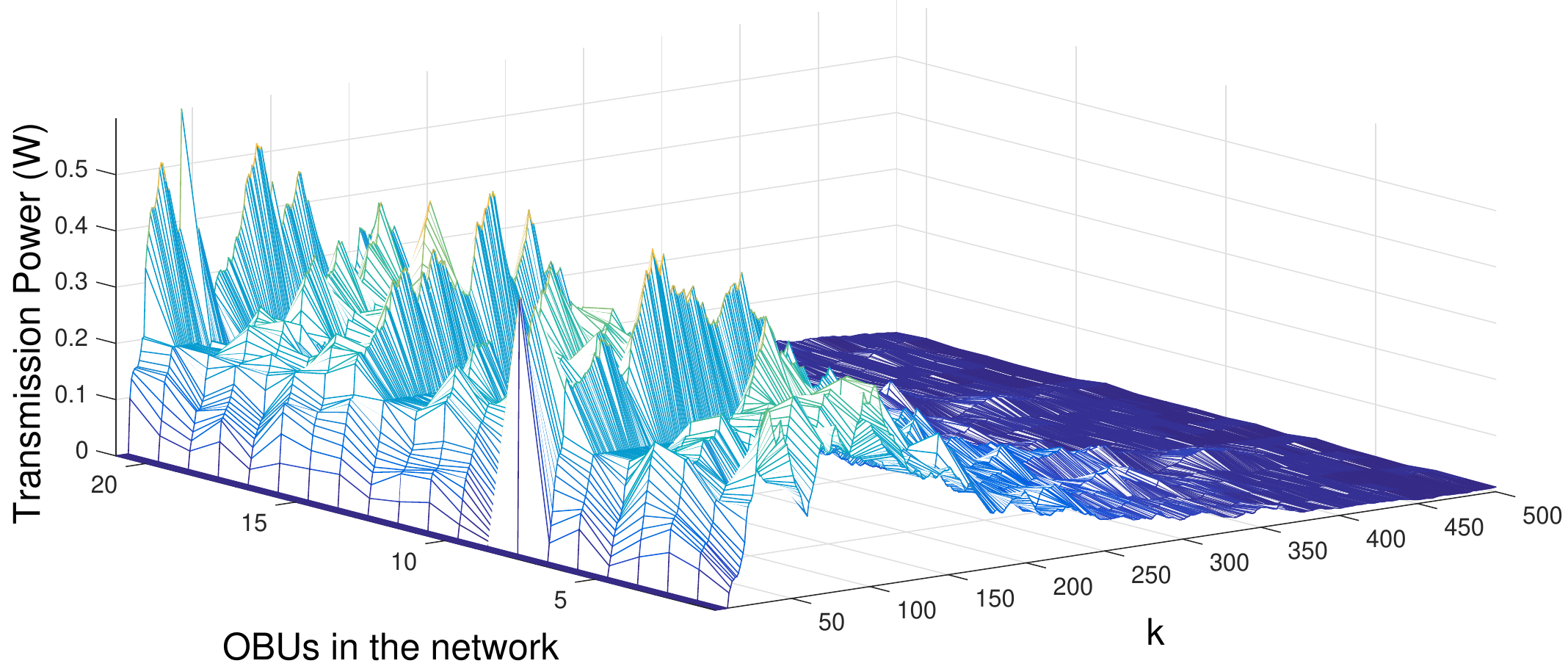}
	\vspace{-0.4cm}
	\caption{ {\textbf{Scenario B}: Sample of the transmission power per OBU in the inner-loops with} speed of 72 km/h (all the OBUs approach the RSU).}
	\label{Power72B}
\end{figure}

Meanwhile, the bottom panel of Fig. \ref{powerSINR72B} shows a realization of the SINR response obtained by the dynamic LQG power allocation algorithm  {in \eqref{LQG} for Scenario B at OBUs constant speeds} of 72 km/h. This plot presents the performance of the 21 OBUs in the VCN, where all the OBUs are able to achieve their objective values despite time-varying mobility profile and interference. On the other hand, the transmission power required to achieve this  {time-varying} objective SINR is illustrated in the top panel of Fig. \ref{powerSINR72B}. Also, Fig. \ref{Power72B} shows a detailed response of the  {transmission} power per OBU in the network  {over the 500 samples of} simulation time, where this parameter keeps decreasing after half the evaluation time, since all OBUs are approaching  {to} the corresponding RSU.

\subsection{Scenario C}

\begin{figure}
	\centering
	\includegraphics[width=1\linewidth]{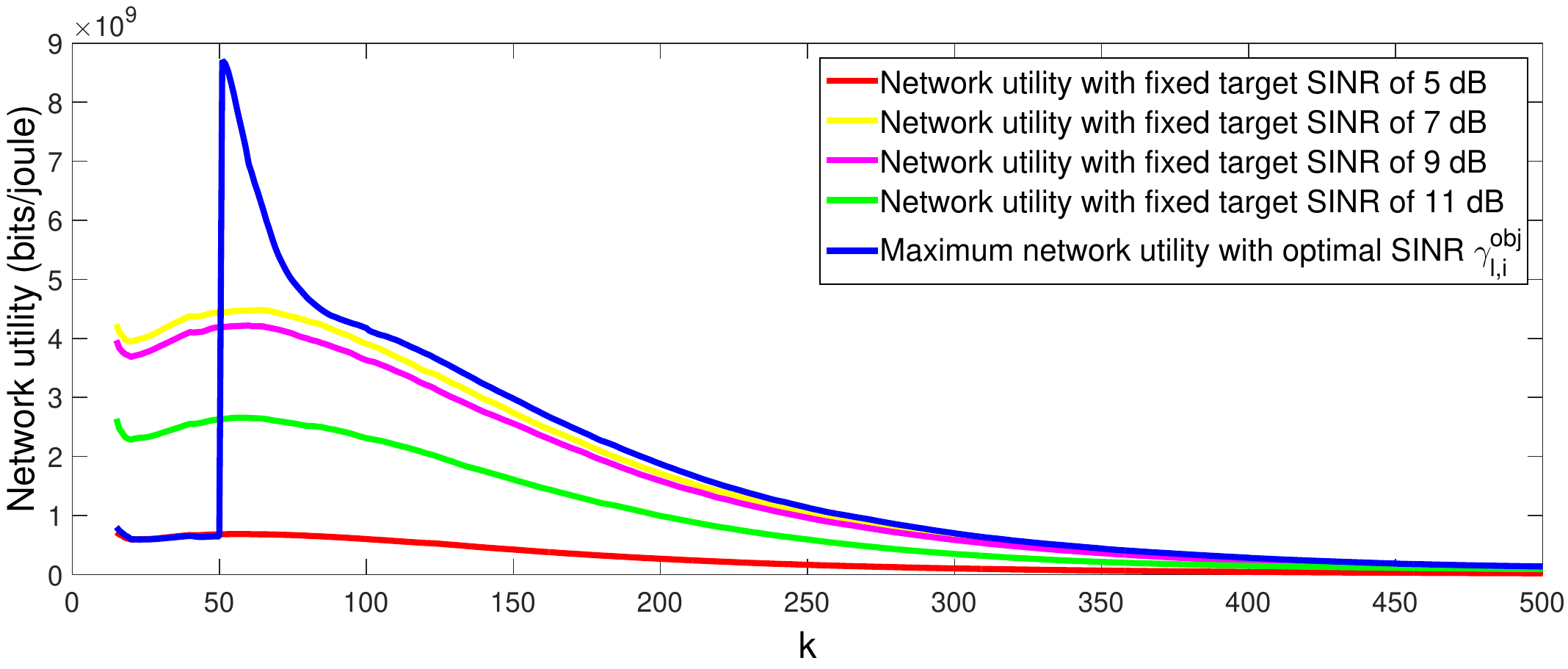}
	\caption{ {\textbf{Scenario C}: Monte Carlo evaluation of instantaneous network utility in \eqref{Uopt}, with 100 closed-loop realization of 500 samples total duration, and OBUs speed of 72 km/h} (all the OBUs move away from the RSU).}
	\label{utilidadredC}
\end{figure}

Figure \ref{utilidadredC} shows the Monte Carlo evaluation with 100 realizations of 500 samples at a speed of 72 km/h, when all the OBUs move away from the RSUs.  Figure \ref{utilidadredC} illustrates that the network utility decreases monotonically during the evaluation time,  {since the propagation channels are reducing their gains by the increased path loss}.  Nonetheless, these results highlight that our proposed scheme once more reaches the maximum utility compared to the fixed objective SINRs  {of 5, 7, 9 and 11 dB} in Fig \ref{DiagUtility}. The abrupt transient in Fig. \ref{utilidadredC} after the initialization time is due to the dynamic effect of the optimization process, and the large step from the fixed objective SINR at 5 dB, during the first 50 samples.

\begin{figure}
	\centering
	\includegraphics[scale=0.22]{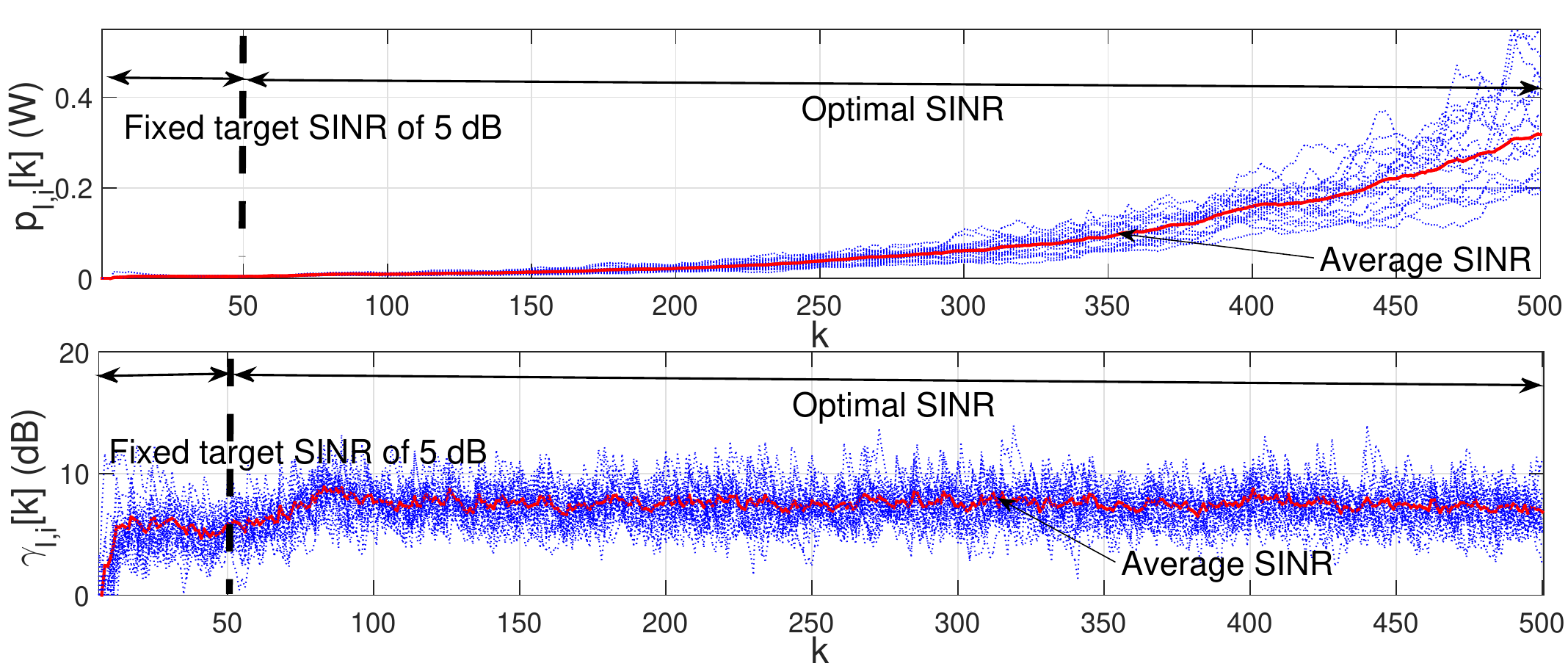}
	\vspace{-0.35cm}
	\caption{ {\textbf{Scenario C}: Sample of the LQG power allocation in the inner-loops with OBUs} speed of 72 km/ h (all the OBUs move away from the RSU): (bottom) instantaneous SINR, and (top) transmission power level.}
	\label{powerSINR72C}
\end{figure}

\begin{figure}
	\centering
	\includegraphics[scale=0.22]{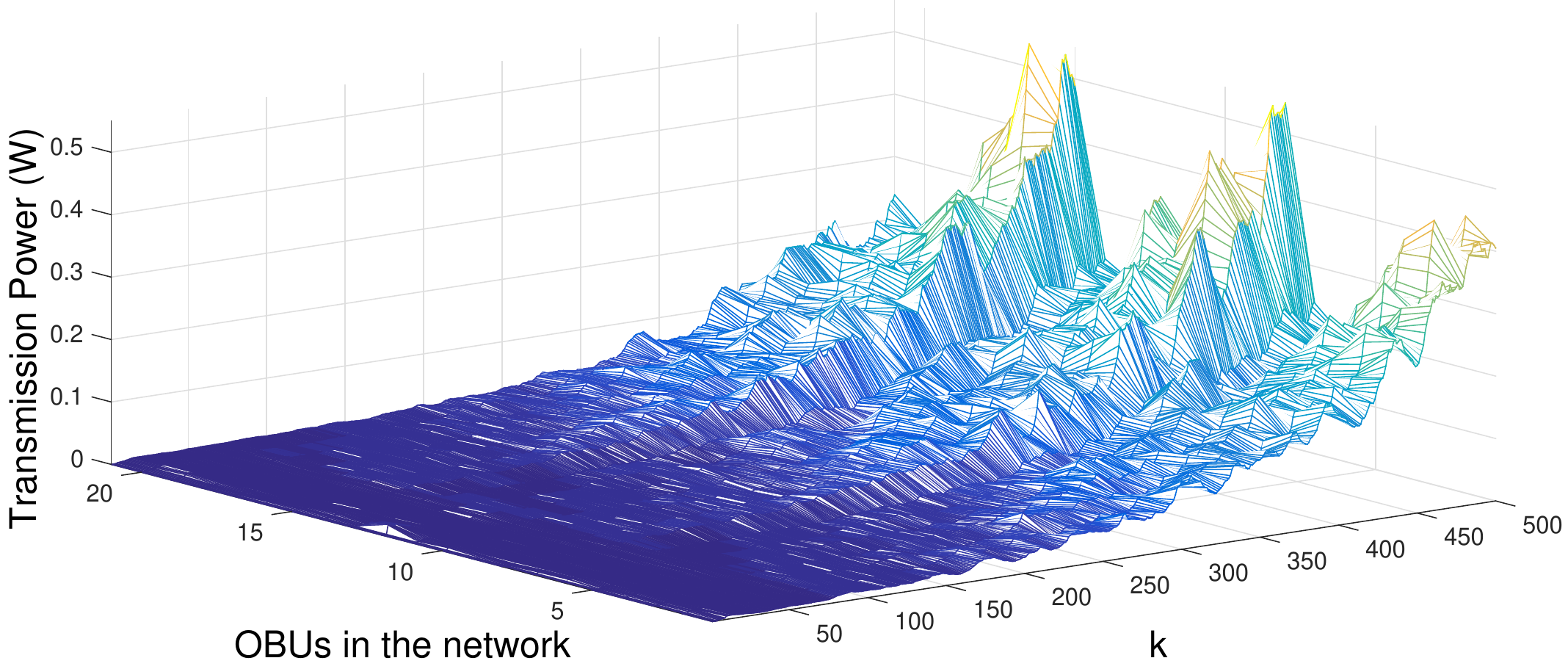}
	\vspace{-0.4cm}
	\caption{ {\textbf{Scenario C}: Sample of the transmission power per OBU in the inner-loops with} speed of 72 km/h (all the OBUs move away from the RSU).}
	\label{Power72C}
\end{figure}

Finally, Fig. \ref{powerSINR72C} shows a realization of the LQG power allocation  {in the inner-loops} over 500 samples in terms of SINR tracking (bottom panel), and required transmission power (top panel). Meanwhile, Fig. \ref{Power72C} shows the resulting transmission power per OBU in the network over  {the 500 samples of the simulation time}. In this scenario, the transmission power has to be constantly raised to maintain the desired objective SINRs, which are defined by the outer-loop (see Fig. \ref{DiagUtility}). Figures \ref{powerSINR72C} and \ref{Power72C} show that the maximum allowable transmission power starting at roughly 450 samples of the simulation time degrades the SINR tracking performance. It is worth mentioning that the performance degradation is caused by the  {OBUs} that operate in channels 5 and 6 of the 5.9 GHz band (short distance channels), since the IEEE 802.11p standard limits their maximum transmission power levels in the uplink and downlink. Nonetheless, as shows Fig. \ref{utilidadredC}, our proposed scheme reached the maximum utility despite this transmission power limitation. 

 {As a general remark, the lowest network utility for all the propagation scenarios described in Figs. \ref{utilidadredA72}, \ref{utilidadredA90}, \ref{utilidadredA108}, \ref{utilidadredB} and \ref{utilidadredC} is obtained with the fixed objective SINR of 5 dB in the inner loops. We highlight that for this low SINR value, the QoS in the V2I communication links achieved a low throughput, despite the low levels of interference, producing the smallest utility function values. Whereas the fixed objective SINR of 11 dB reached the second lowest network utility. But In this latter case, the high objective SINR required high transmission power of the OBUs which produced larger interference, and also a larger weight in the denominator of \eqref{uil}, so the resulting utility was also low. Consequently, from the results in Figs. \ref{utilidadredA72}, \ref{utilidadredA90}, \ref{utilidadredA108}, \ref{utilidadredB} and \ref{utilidadredC}, time-varying objective SINRs between 7 and 9 dB are needed by the inner-loops to achieve the maximum network utility, which are supplied by the outer-loop in our proposal of Fig. \ref{DiagUtility}.}
 
  {To demonstrate the practicability of our network utility optimization scheme, the simulation time cost was computed during the Monte Carlo evaluations for all propagation scenarios with a speed of 72 km/h. The computing platform had the following technical characteristics: Intel$^R$ Core$^{TM}$ i3-6100U CPU @2.3 GHz and 16 GB RAM. The mean computational time$\pm$standard deviation for the three conditions were: 9.79$\pm$1.83 s (\textbf{Scenario A}),  9.98$\pm$1.71 s (\textbf{Scenario B}), and 9.96$\pm$1.45 s (\textbf{Scenario C}). Hence, there is no significant difference in the mean-time cost for the three mobility scenarios with a similar variability. Moreover, our results highlight a low time cost for our application of network utility optimization.}

\section{Conclusion}

In this paper, we proposed to maximize network utility based on the IEEE 802.11p standard, where the uplink of a vehicular communication system (vehicle-to-infrastructure) was addressed. First, the IEEE 802.11p standard was studied to analyze the relation among interference, SINR and transmission power in each OBU of the network. The problem of utility maximization in a VCN was formulated mathematically, and the optimality conditions were derived.  Based on the relation between transmission power and SINR due to interference in the network, a  {distributed} solution was proposed for utility maximization,  {which is implemented at three hierarchical levels: network central unit, RSUs, and OBUs. The proposed solution involved a two-loop feedback structure at different time-scales: the network central unit coordinates the outer-loop that computes the optimal objective SINRs, according to the degree of interference in the network and to maximize network utility; meanwhile the inner-control loops involve the RSU and OBUs feedback paths to update the transmission power dynamically with the aim of compensating channel variations and network latency.}  Our results showed that the proposed scheme not only guaranteed the maximum utility of the network, but also improved its power efficiency which allows to reduce the interference between adjacent channels, and the interference from RSU to RSU. Finally, our numerical evaluation illustrated that the proposed scheme obtained the optimum SINRs which achieved maximum network utility compared to  fixed  {objective} SINRs (i.e. 5 dB, 7 dB, 9 dB and 11 dB). Furthermore, the results of the evaluation highlighted that the maximum utility is consistent for the diverse testing scenarios and mobility profiles  {with a low time cost}.  {The analytical derivations and validation stage of our network utility maximization scheme leave many important directions of future work: (i) derive a downlink implementation of the distributed maximization scheme, (ii) formulate the model and corresponding solution to the LTE-V standard, and (iii) carry out an empirical evaluation of the proposed distributed optimization scheme.} 

\section*{Acknowledgment}

The authors thank CONACYT for supporting Miguel A. Diaz-Ibarra with  a doctoral fellowship, and by a Basic Science Grant (Ref. \# 254637)

\section*{Appendix}

 From the utility definition in (\ref{UtiDef}) for the $ i $-th OBU at $ l $-th RSU, and taking partial derivatives \cite{Differentiation}, we obtain (for simplicity the dependence on the time index $k$ will be omitted):
 \begin{align}
\frac{\partial u_{l,i}}{\partial p_{l,i}} & = \dfrac{w_{l,i}}{(p_{l,i})^{2}}\left\lbrace p_{l,i} f'(\gamma_{l,i})\frac{\partial \gamma_{l,i}}{\partial p_{l,i}}- f(\gamma_{l,i}) \right\rbrace~, \label{uli} \\
\frac{\partial u_{l,i}}{\partial p_{m,i}} & = \dfrac{w_{l,i}}{p_{l,i}} f'(\gamma_{l,i})\frac{\partial \gamma_{l,i}}{\partial p_{m,i}}~, \label{umi} \\
\frac{\partial u_{l,i}}{\partial p_{l,j}} & = \dfrac{w_{l,i}}{p_{l,i}} f'(\gamma_{l,i})\frac{\partial \gamma_{l,i}}{\partial p_{l,j}}~. \label{ulj}
\end{align}
Next, the partial derivatives of the SINR $\gamma_{l,i}$ with respect to the transmission power components are computed:
\begin{equation} 
\frac{\partial \gamma_{l,i}}{\partial p_{l,i}} = H_{l,i}^{l}~,  \quad
\frac{\partial \gamma_{l,i}}{\partial p_{m,i}}  = -\gamma_{l,i} H_{l,i}^{m}~, \quad 
\frac{\partial \gamma_{l,i}}{\partial p_{l,j}}  =- \gamma_{l,i} H_{l,i}^{j}~, \label{gammal}
\end{equation}
where
 \begin{align}
H_{l,i}^{l} & \triangleq \frac{\dfrac{W}{r_{l,i}} | h_{l,i}| ^{2} }{  I_{l,i}  +   \sum_{h=1,h\neq l}^{M}p_{h,i} |h_{h,i}|^{2}  +\sigma^{2}_{l,i}}~,\\  
H_{l,i}^{m} & \triangleq  \frac{ | h_{m,i}| ^{2} }{  I_{l,i}  +   \sum_{h=1,h\neq m}^{M}p_{h,i} |h_{h,i}|^{2}  +\sigma^{2}_{l,i}}~,\\  
H_{l,i}^{j} & \triangleq \frac{ \frac{\partial I_{l,i}}{\partial p_{l,j}}}{  I_{l,i}  +   \sum_{h=1,h\neq l}^{M}p_{h,i} |h_{h,i}|^{2}  +\sigma^{2}_{l,i}}~.
\end{align}
Then, the partial derivative of the ACI $I_{l,i}$ with respect to each transmission power component is calculated as:
 \begin{equation}
	\frac{\partial I_{l,i}}{\partial p_{l,j}} = 
	\left \{
	\begin{array}{cl}
	\sum_{j=i-1, j\neq i}^{i+1} c_{l,j}|h_{l,j}|^{2} &\textrm{if} \; \;  j \in [i-1,i+1]~, \\
	&  \forall i \in \{2,...,U-1\} \;   \\
	c_{l,j}|h_{l,j}|^{2} & \textrm{if} \; \; i=1 \; \Rightarrow \;  j=2 \; \;  \vee\\
	&    i=U-1 \;  \Rightarrow \; j=U \; \\ 
	0 &    \mbox{otherwise}
	\end{array}
	\right . ~,
\end{equation}
where the interference coefficients $\{c_{l,j}\}$ are defined in \eqref{interf}.  By a direct substitution of (\ref{gammal})  into (\ref{uli}), (\ref{umi}) and (\ref{ulj}), respectively, it is deduced that:
\begin{align}
	\frac{\partial u_{l,i}}{\partial p_{l,i}} & = \dfrac{w_{l,i}}{(p_{l,i})^{2}}\left\lbrace  f'(\gamma_{l,i}) \underbrace{p_{l,i} H_{l,i}^{l}}_{\gamma_{l,i}}   - f(\gamma_{l,i}) \right\rbrace~,\\
	 & =\dfrac{w_{l,i}}{(p_{l,i})^{2}}\left\lbrace  f'(\gamma_{l,i})\gamma_{l,i}- f(\gamma_{l,i}) \right\rbrace~, \label{ulis} \\
	\frac{\partial u_{l,i}}{\partial p_{m,i}} & =-\dfrac{w_{l,i}}{p_{l,i}} f'(\gamma_{l,i}) \gamma_{l,i} H_{l,i}^{m}~, \label{umis} \\
	\frac{\partial u_{l,i}}{\partial p_{l,j}} & =\left \{
	\begin{array}{cl}
	- \dfrac{w_{l,i}}{p_{l,i}} f'(\gamma_{l,i}) \gamma_{l,i} H_{l,i}^{j} & \textrm{if} \; \; j \in [i-1,i+1]~, \\ &  \forall i \in \{2,...,U-1\} \;   \\
	- \dfrac{w_{l,i}}{p_{l,i}} f'(\gamma_{l,i}) \gamma_{l,i} H_{l,i}^{j} & \textrm{if} \;  \;  i=1 \; \Rightarrow \;  j=2  \; \; \vee\\
	&   i=U-1 \; \Rightarrow  \; j=U  \\ 
	0 &   \mbox{otherwise}~.
	\end{array} \right . \label{uljs}
\end{align}
The partial derivative of the network utility $\mathcal{U}$ with respect to the transmission power for the $ i $-th OBU to $ l $-th RSU is:
\begin{equation}
	\frac{\partial \mathcal{U}(\mathbf{p}_{1}, \ldots , \mathbf{p}_{M})}{\partial p_{l,i}}= \sum_{h=1}^{M} \sum_{j=1}^{U} \frac{\partial u_{h,j}}{\partial p_{l,i}}~.
	\label{U}
\end{equation}
By a direct substitution of (\ref{ulis}), (\ref{umis}) and (\ref{uljs})
into (\ref{U}), it is obtained:
\begin{align}
	\frac{\partial \mathcal{U}(\mathbf{p}_{1}, \ldots , \mathbf{p}_{M})}{\partial p_{l,i}} & = - \sum_{m=1,m\neq l}^{M}\dfrac{w_{l,i}}{p_{l,i}}f(\gamma_{m,i})H_{m,i}^{l}\gamma_{m,i} \nonumber\\
	& \qquad - \; \sum_{m=1,m\neq l}^{M}\dfrac{w_{l,j}}{p_{l,j}}f(\gamma_{l,j})H_{l,j}^{i}\gamma_{l,j} \\ 
	& \qquad + \; \dfrac{w_{l,i}}{(p_{l,i})^{2}}\left\lbrace  f'(\gamma_{l,i})\gamma_{l,i}- f(\gamma_{l,i}) \right\rbrace ~. \nonumber
\end{align}
As a result, if the partial derivative of the network utility in \eqref{U} is set equal to zero, it is deduced an algebraic condition for optimality which is crucial in our distributed strategy:
\begin{equation}  
f(\gamma_{l,i})\gamma_{l,i}-f(\gamma_{l,i})=  \hat{M}_{l,i}(\boldsymbol{\gamma}, \mathbf{p})\; \; \forall i \in [1,U], \; l \in [1,M]~,   \label{eq24}
\end{equation}
where
\begin{align}
\hat{M}_{l,i}(\boldsymbol{\gamma}, \mathbf{p}) & \triangleq \dfrac{(p_{l,i})^{2}}{w_{l,i}}  \left \{  \sum_{m=1,m\neq l}^{M}\dfrac{w_{l,i}}{p_{l,i}}f(\gamma_{m,i})H_{m,i}^{l}\gamma_{m,i} \right .   \nonumber\\
& \qquad  \left . + \sum_{j \in J_{i}}^{}\dfrac{w_{l,j}}{p_{l,j}}f(\gamma_{l,j})H_{l,j}^{i}\gamma_{l,j} \right \} ~, \label{Mli} \\
\boldsymbol{\gamma} & =[\gamma_{1,1} \; \ldots \; \gamma_{M,U}]^\top ~, \\
\mathbf{p} & =[p_{1,1} \; \ldots \; p_{M,U}]^\top~, 
\end{align}
and
\begin{equation}
	J_{i} = 
	\left \{
	\begin{array}{cl}
	[i-1,i+1] & \textrm{if} \; \;  i \in \{2,...,U-1\} \;   \\
	2 & \textrm{if} \; \; i=1  \; \\ 
	U-1 & \textrm{if} \; \;  i=U
	\end{array}
	\right . ~.
\end{equation}
With respect to (\ref{Mli}), we have a non-negative property $\hat{M}_{l,i} \geq 0$ of this variable, since this term agglomerates interference with other RSUs in the network, and between OBUs with adjacent channels. In fact, the optimality conditions in \eqref{eq24} involve $M \cdot U$ coupled nonlinear algebraic equations that have to be solved at each time instant. Furthermore, according to \cite {PreEqualization}, the highest value of $\gamma_{l, i} $ that satisfies \eqref{eq24} ($\gamma_{l, i}^{max}$) is obtained when $\hat{M}_{l,i}=0$ (no-interference condition). As a result, for $\hat{M}_{l,i} > 0$, the optimal $\gamma_{l, i} $ will be lower to $\gamma_{l, i}^{max}$ due to interference. Another relevant property of \eqref{eq24} is that each OBU will have a different optimal SINR $\gamma_{l,i}$ to maximize the network utility.

\ifCLASSOPTIONcaptionsoff
  \newpage
\fi

\bibliographystyle{IEEEtran}
\bibliography{IEEEabrv,./PaperTVT.bib}

\end{document}